\documentclass[aps,superscriptaddress,floatfix,byrevtex,preprint,amsmath,amssymb]{revtex4} 
\usepackage{epsfig}

\usepackage{graphicx}               
\usepackage{dcolumn}                
\usepackage{multirow}               
\usepackage{amsmath}                
\usepackage{color}
\usepackage{comment}
\graphicspath{{ps}}                 

\newcommand{\mb}{{M_{\rm bc}}}
\newcommand{\de}{{\Delta{E}}}
\newcommand{\bp}{B^{+}}

\newcommand{\bz}{B^{0}}

\newcommand{\pp}{{p\bar{p}}}
\newcommand{\ppk}{{p\bar{p}K^+}}
\newcommand{\pppi}{{p\bar{p}\pi^+}}

\newcommand{\ppkst}{{p\bar{p}K^{* +}}}
\newcommand{\ppkstz}{{p\bar{p}K^{*0}}}

\newcommand{\mll}{M_{\Lambda\bar{\Lambda}}}
\newcommand{\LL}{\Lambda\bar{\Lambda}}
\newcommand{\llk}{{\Lambda\bar{\Lambda}K^+}}
\newcommand{\llpi}{{\Lambda\bar{\Lambda}\pi^+}}
\newcommand{\llkz}{{\Lambda\bar{\Lambda}K^0}}

\newcommand{\llkst}{{\Lambda\bar{\Lambda}K^{* +}}}
\newcommand{\llkstz}{{\Lambda\bar{\Lambda}K^{*0}}}

\newcommand{\plpi}{{p\bar{\Lambda}\pi^-}}

\newcommand{\kst}{{K^{*{+}}}}
\newcommand{\kstz}{{K^{*0}}}
\newcommand{\ks}{{K_S^0}}
\newcommand{\kz}{{K^0}}

\begin{document}

\preprint{\vbox{ \hbox{   }
                 \hbox{Belle Preprint 2008-30}
                 \hbox{KEK Preprint 2008-41}
}}

\title{ \quad\\[0.5cm] Observation of $\bz \to \llkz$ and $\bz \to \llkstz$ at Belle}

\affiliation{Budker Institute of Nuclear Physics, Novosibirsk}
\affiliation{Chiba University, Chiba}
\affiliation{University of Cincinnati, Cincinnati, Ohio 45221}
\affiliation{T. Ko\'{s}ciuszko Cracow University of Technology, Krakow}
\affiliation{The Graduate University for Advanced Studies, Hayama}
\affiliation{Hanyang University, Seoul}
\affiliation{University of Hawaii, Honolulu, Hawaii 96822}
\affiliation{High Energy Accelerator Research Organization (KEK), Tsukuba}
\affiliation{Hiroshima Institute of Technology, Hiroshima}
\affiliation{Institute of High Energy Physics, Chinese Academy of Sciences, Beijing}
\affiliation{Institute of High Energy Physics, Vienna}
\affiliation{Institute of High Energy Physics, Protvino}
\affiliation{Institute for Theoretical and Experimental Physics, Moscow}
\affiliation{J. Stefan Institute, Ljubljana}
\affiliation{Kanagawa University, Yokohama}
\affiliation{Korea University, Seoul}
\affiliation{Kyungpook National University, Taegu}
\affiliation{\'Ecole Polytechnique F\'ed\'erale de Lausanne (EPFL), Lausanne}
\affiliation{Faculty of Mathematics and Physics, University of Ljubljana, Ljubljana}
\affiliation{University of Maribor, Maribor}
\affiliation{University of Melbourne, School of Physics, Victoria 3010}
\affiliation{Nagoya University, Nagoya}
\affiliation{Nara Women's University, Nara}
\affiliation{National Central University, Chung-li}
\affiliation{National United University, Miao Li}
\affiliation{Department of Physics, National Taiwan University, Taipei}
\affiliation{H. Niewodniczanski Institute of Nuclear Physics, Krakow}
\affiliation{Nippon Dental University, Niigata}
\affiliation{Niigata University, Niigata}
\affiliation{University of Nova Gorica, Nova Gorica}
\affiliation{Osaka City University, Osaka}
\affiliation{Osaka University, Osaka}
\affiliation{Panjab University, Chandigarh}
\affiliation{Saga University, Saga}
\affiliation{University of Science and Technology of China, Hefei}
\affiliation{Seoul National University, Seoul}
\affiliation{Sungkyunkwan University, Suwon}
\affiliation{University of Sydney, Sydney, New South Wales}
\affiliation{Tata Institute of Fundamental Research, Mumbai}
\affiliation{Toho University, Funabashi}
\affiliation{Tohoku Gakuin University, Tagajo}
\affiliation{Tohoku University, Sendai}
\affiliation{Department of Physics, University of Tokyo, Tokyo}
\affiliation{Tokyo University of Agriculture and Technology, Tokyo}
\affiliation{IPNAS, Virginia Polytechnic Institute and State University, Blacksburg, Virginia 24061}
\affiliation{Yonsei University, Seoul}
  \author{Y.-W.~Chang}\affiliation{Department of Physics, National Taiwan University, Taipei} 
  \author{M.-Z.~Wang}\affiliation{Department of Physics, National Taiwan University, Taipei} 
  \author{I.~Adachi}\affiliation{High Energy Accelerator Research Organization (KEK), Tsukuba} 
  \author{H.~Aihara}\affiliation{Department of Physics, University of Tokyo, Tokyo} 
  \author{T.~Aushev}\affiliation{\'Ecole Polytechnique F\'ed\'erale de Lausanne (EPFL), Lausanne}\affiliation{Institute for Theoretical and Experimental Physics, Moscow} 
  \author{A.~M.~Bakich}\affiliation{University of Sydney, Sydney, New South Wales} 
  \author{V.~Balagura}\affiliation{Institute for Theoretical and Experimental Physics, Moscow} 
  \author{A.~Bay}\affiliation{\'Ecole Polytechnique F\'ed\'erale de Lausanne (EPFL), Lausanne} 
  \author{V.~Bhardwaj}\affiliation{Panjab University, Chandigarh} 
  \author{U.~Bitenc}\affiliation{J. Stefan Institute, Ljubljana} 
% ...
\collaboration{The Belle Collaboration} 

\noaffiliation

 \begin{abstract} 

  We study the charmless decays $B \to \LL h$, where $h$ stands for $\pi^+$, $K^+$,
  $K^0$,$K^{*+}$, or $K^{*0}$, 
  using a $605\,{\rm fb}^{-1}$ data sample collected at the $\Upsilon(4S)$ resonance
  with the Belle detector at the KEKB asymmetric energy $e^+ e^-$ collider. 
  We observe $\bz \to \llkz$ and $\bz \to \llkstz$ with branching fractions 
  of $(4.76^{+0.84}_{-0.68} (stat.) \pm 0.61 (syst.)) \times 10^{-6}$ and
  $(2.46^{+0.87}_{-0.72} \pm 0.34) \times 10^{-6}$, respectively.
  The significances of these signals
{\color{black} 
  in the threshold-mass enhanced mass region are $12.4\sigma$ and $9.3\sigma$, respectively.
}
 We also update the branching fraction $\mathcal{B} (\bp \to \llk) = (3.38^{+0.41}_{-0.36}
  \pm 0.41 )\times 10^{-6}$ with better accuracy,
  and report the following measurement or 90\% confidence level upper limit
  in the threshold-mass-enhanced region:
  $\mathcal{B} (\bp \to \llkst) = (2.19^{+1.13}_{-0.88}
  \pm 0.33)  \times 10^{-6}$ with
  3.7$\sigma$ significance; 
  $\mathcal{B} (\bp \to \llpi) < 0.94 \times 10^{-6}$.
  A related  search for $\bz \to \LL\bar{D}^0$ yields a branching fraction
  $\mathcal{B} (\bz \to \LL\bar{D}^0 ) = (1.05_{-0.44}^{+0.57} \pm 0.14)\times 10^{-5}$.
  This may be compared with the large, $\sim 10^{-4}$, branching
  fraction observed for $\bz \to \pp\bar{D^0}$.
  The $\mll$ enhancements near
  threshold and related angular distributions for the observed modes
  are also reported.



 \end{abstract}  
\maketitle


\renewcommand{\thefootnote}{\fnsymbol{footnote}}
\setcounter{footnote}{0}
\clearpage

\section{Introduction}
The $b \to s$ penguin loop process plays an important role
in rare $B$ meson decays~\cite{MEPeskin}. 
It could be sensitive to new physics beyond {\color{black}the} standard
model due to additional contributions from as yet-unknown 
heavy virtual particles in the loop.  
Recently the study of the penguin dominated baryonic $B$ decays
$\bp \to \ppk$~\cite{Wei} and $\bz \to \plpi$~\cite{Wang} 
gave intriguing results.
The proton polar angular distributions in the baryon-antibaryon helicity frame
disagree with the expectations for short distance
$b \to s$ weak decays~\cite{Suzuki07}. 
However, in $B \to \pp K^*$ decays~\cite{JHChen}, 
the $\kstz$ seems to be fully polarized in the helicity zero state 
{\color{black} in agreement} with the $b \to s$ weak decay hypothesis.
The theoretical hierarchies, $\mathcal{B}(\bp \to \ppk) > \mathcal{B}(\bp \to \ppkst)$ 
and $\mathcal{B}(\bp \to \ppkst) > \mathcal{B}(\bz \to \ppkstz)$, from the pole model~\cite{HYCheng}
are experimentally established
{\color{black}
although the predicted branching fraction $\mathcal{B}(\bz \to \ppkstz)$ 
is about a factor of 20 smaller than the experimental measurement.
}
It is therefore interesting to study the corresponding branching fractions for $B \to \LL K^{(*)}$ decays, 
the counterparts with protons replaced by $\Lambda$'s.

\begin{figure}[htb]
\begin{center}
\hskip -7.4cm {\bf (a)} \hskip 6.6cm {\bf (b)}\\
\vskip -1.2cm
\includegraphics[width=0.43\textwidth]{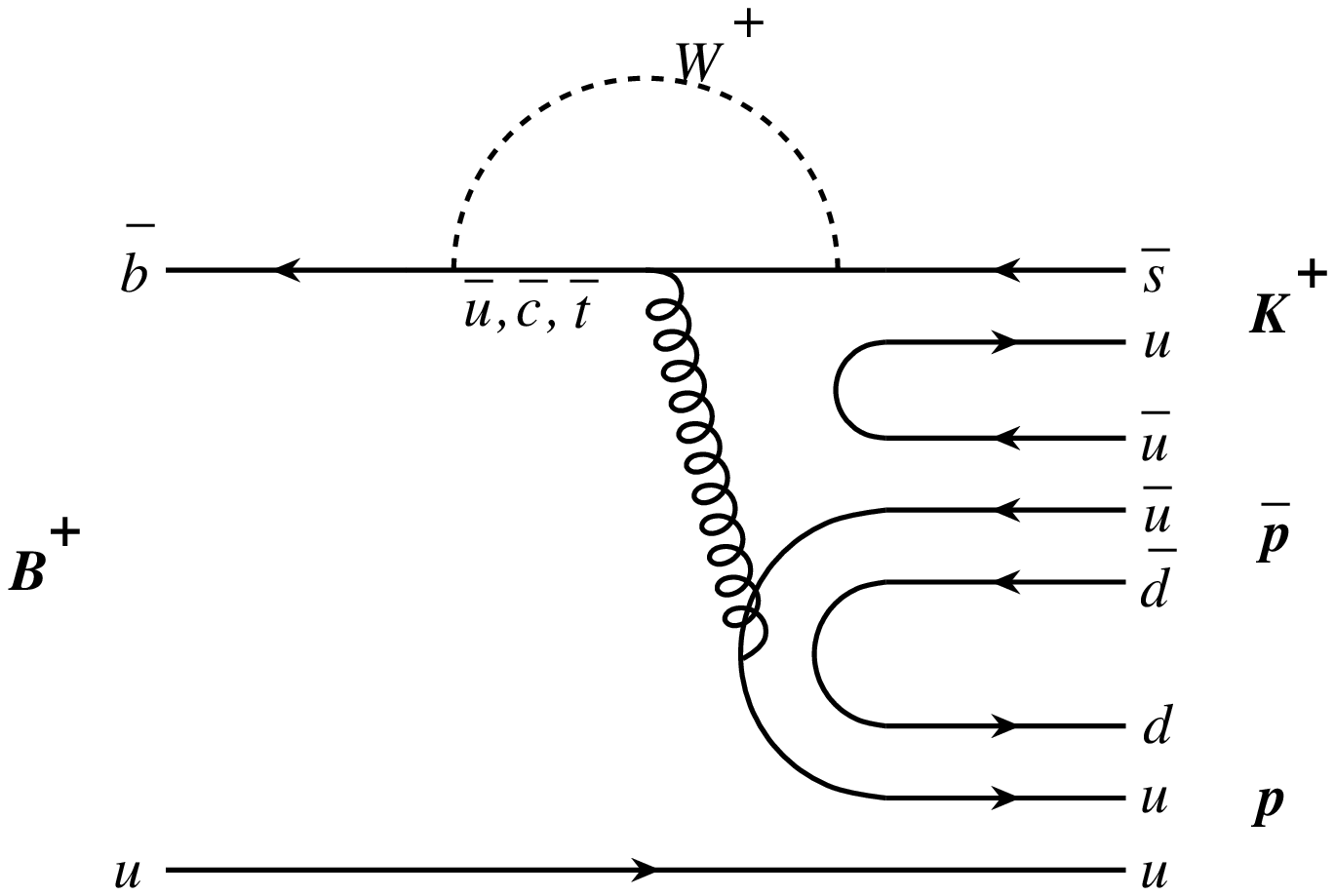}
\includegraphics[width=0.43\textwidth]{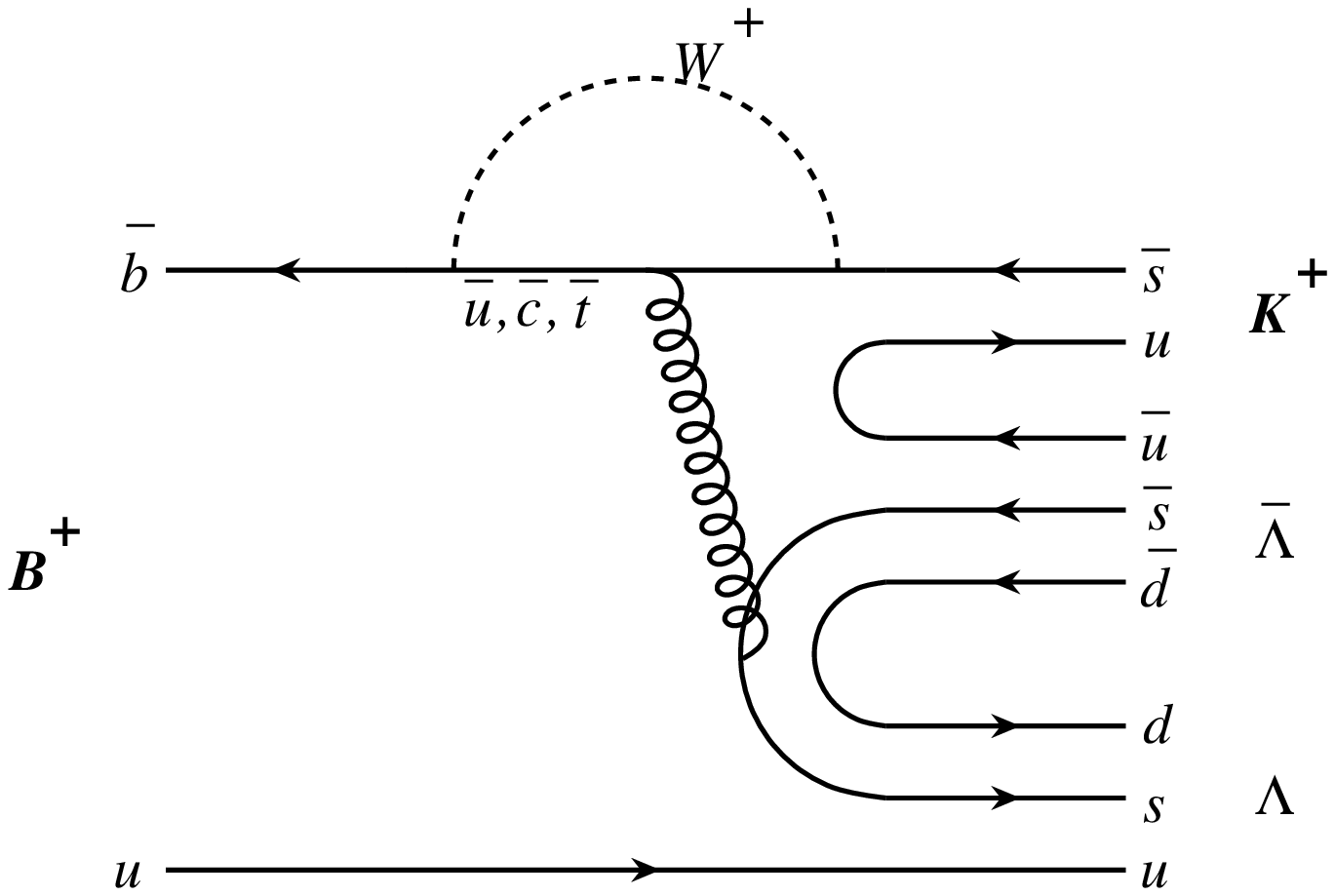}\\
\hskip -7.4cm {\bf (c)} \hskip 6.6cm {\bf (d)}\\
\vskip -1.2cm
\includegraphics[width=0.43\textwidth]{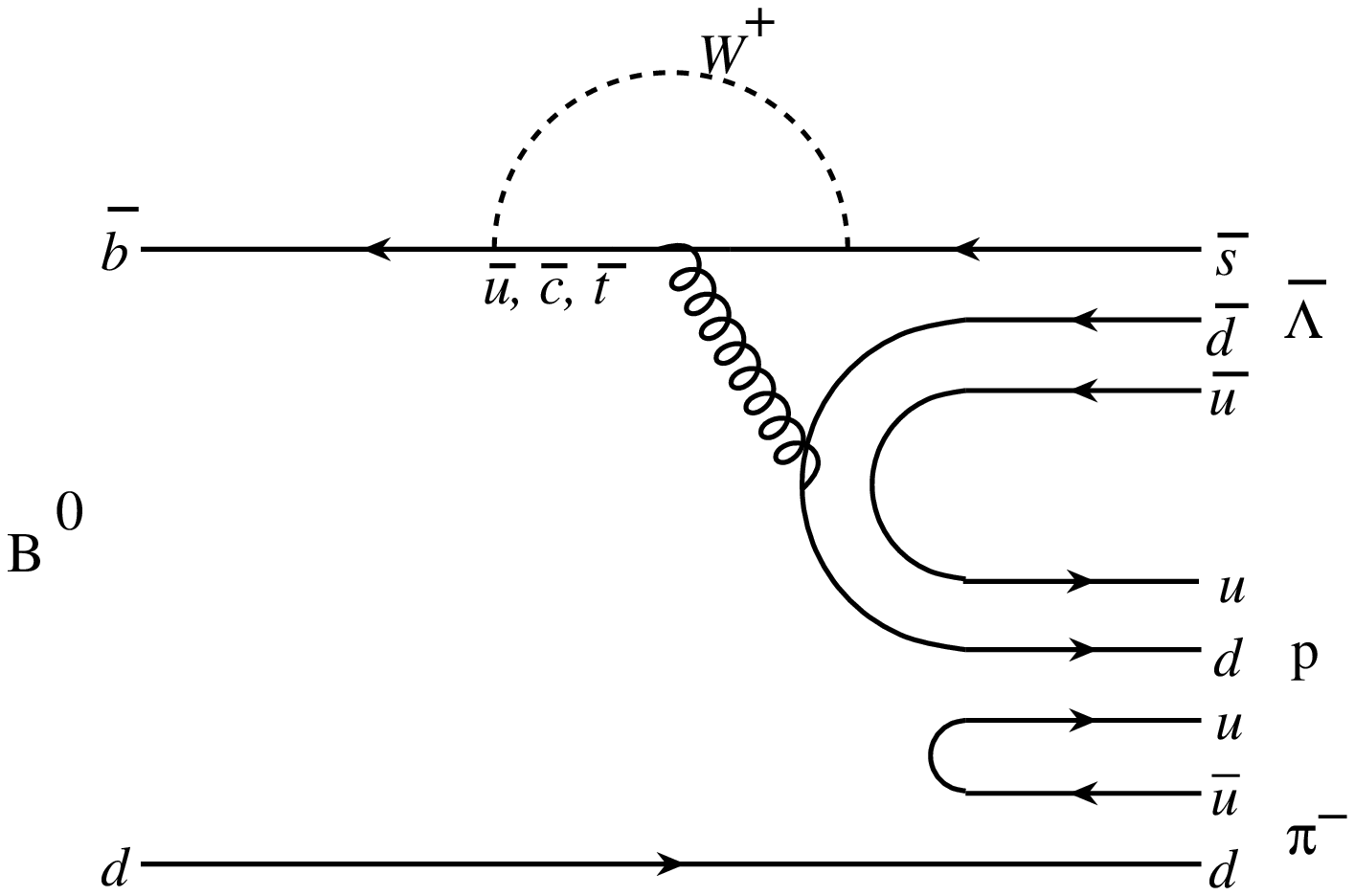}
\includegraphics[width=0.43\textwidth]{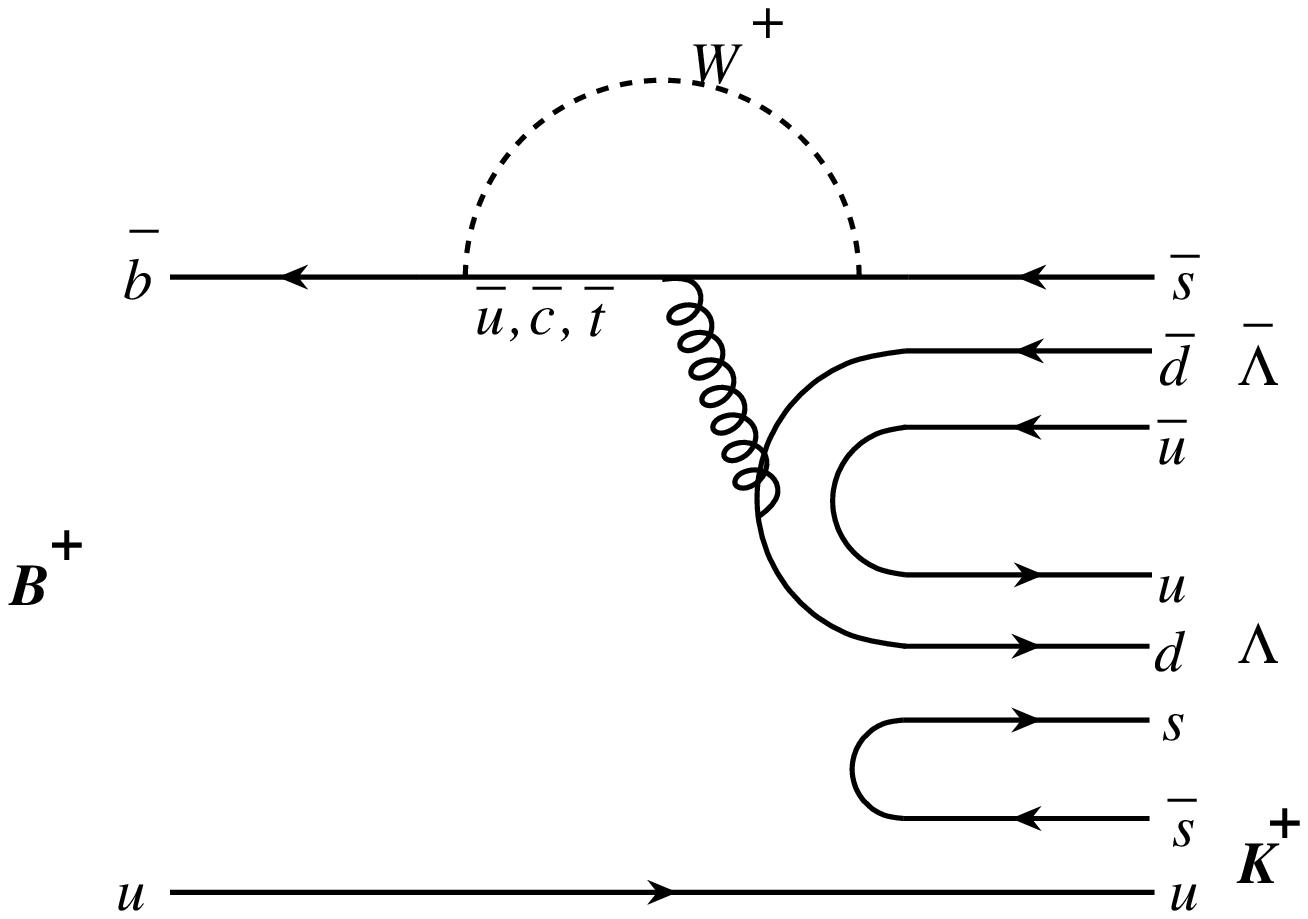}\\
\caption{Comparisons of possible decay diagrams between $\bp \to \ppk$/$\bz \to \plpi$ and 
 $\bp \to \llk$.}
\label{fg:feyn}
\end{center}
\end{figure}

In this paper, we study the charmless three-body decays $B \to \LL h$, 
where $h$ stands for $\pi^+$, $K^+$, $K^0$, $K^{*+}$, or $K^{*0}$~\cite{conjugate}. 
The mode $\bp \to \llk$ has been previously observed~\cite{YJLee} and 
presumably proceeds through a $\bar{b} \to \bar{s}s\bar{s}$ 
process. 
This decay process can be related to $\bp \to \ppk$ as shown 
in Fig.~\ref{fg:feyn}(a) and Fig.~\ref{fg:feyn}(b). 
One can simply replace the $ud - \bar{u}\bar{d}$ diquark pair
with an $sd - \bar{s}\bar{d}$ pair to establish a one-to-one correspondence 
between $B \to \pp h$ and $\LL h$ decays. 
A common feature of these decays is that the baryon-antibaryon mass
spectra peak near threshold as conjectured in Refs.~\cite{Suzuki07, HS}.
The $K^+$ meson carries the energetic $\bar{s}$ quark 
from the $\bar{b} \to \bar{s}$ 
transition so that a threshold enhancement of the baryon and
antibaryon system is naturally formed. 
However, there is another possibility
shown 
in Fig.~\ref{fg:feyn}(c) and Fig.~\ref{fg:feyn}(d), 
where the $\bar{\Lambda}$ (instead of the $K^+$) 
carries the $\bar{s}$ from the $\bar{b} \to \bar{s}$ transition. 
It is interesting to know the role of this $\bar{s}$ quark in $B \to \LL K^{(*)}$ weak decays.

Since the branching fractions of $B \to \LL K$ and $\llpi$ decays 
are theoretically expected at a level~\cite{prediction} {\color{black}that} is 
detectable with our present data sample, 
we attempt to determine the
branching fractions of the various $B \to \LL h$ decays and compare
with the latest measurements for $B \to \pp h$. 
We also examine the low mass $\mll$ enhancements near
threshold and the related angular distributions in order to investigate the underlying dynamics. 

\section{Event Selection and Reconstruction}
\subsection{Data Samples and the Belle Detector}
For this study,
we use a  605 fb$^{-1}$  data sample, consisting of 657 $ \times 10^6 B\bar{B}$ pairs,
collected with the Belle detector on the $\Upsilon({\rm 4S})$ resonance
at the KEKB asymmetric energy $e^+e^-$ (3.5 and 8~GeV) collider~\cite{KEKB}.
The Belle detector is a large-solid-angle magnetic spectrometer 
that consists of a silicon vertex detector (SVD),
a 50-layer central drift chamber (CDC), 
an array of aerogel threshold Cherenkov counters (ACC),
a barrel-like arrangement of time-of-flight scintillation counters (TOF), 
and an electromagnetic calorimeter composed of CsI(Tl) crystals 
located inside a super-conducting solenoid coil that provides a 1.5~T magnetic field. 
An iron flux-return located outside of the coil 
is instrumented to detect $K_L^0$ mesons and to identify muons. 
The detector is described in detail elsewhere~\cite{Belle}.

\subsection{Selection Criteria}
The event selection criteria are based on information obtained
from the tracking system (SVD and CDC) and 
the hadron identification system (CDC, ACC, and TOF).
All charged tracks not associated with long lived particles
are required to satisfy track quality criteria
based on track impact parameters relative
to the interaction point (IP). 
The deviations of charged tracks from the IP position are required to be within
$\pm$0.3 cm in the transverse ($x$--$y$) plane, and within $\pm$3 cm
in the $z$ direction, where the $z$ axis is defined to be the direction opposite to the
positron beam.
For each track, the likelihood values $L_p$,
$L_K$, and $L_\pi$ for the proton, kaon, or pion hypotheses, respectively,
are determined from the information provided by
the hadron identification system. A track is identified 
as a kaon if
$L_K/(L_K+L_{\pi})> 0.6$, or as a pion if $L_{\pi}/(L_K+L_{\pi})> 0.6$.
This selection is about 86\% (93\%) efficient for kaons (pions)
while removing about 96\% (94\%) of pions (kaons).
$\ks$ candidates are reconstructed from pairs of oppositely charged tracks 
(both treated as pions) with an invariant mass in the range 
$ 485$ MeV/c$^2 < M_{\pi^+\pi^-} < 510 $ MeV/c$^2$.
The dipion candidate must have a displaced vertex and flight 
direction consistent with a $\ks$ originating from the interaction point. 
We use the selected kaons and pions to form $\kst$
($\to \ks\pi^+$) and  $\kstz$ ($\to K^+\pi^-$) candidates.
Events with a $K^*$ candidate mass between  0.6 GeV/c$^2$ and  1.2 GeV/c$^2$ 
are used for further analysis. 
Similarly, we select $\Lambda$ baryons 
by applying the {\color{black} $\ks$ vertex displacement and flight direction} 
selection criteria to pairs of oppositely
charged tracks---treated as a proton and negative pion---whose  mass
is consistent with the nominal $\Lambda$ baryon mass,
1.111 GeV/c$^2 < M_{p\pi^-} <1.121 $ GeV/c$^2$ {\color{black} \cite{lambdamass}}.
The proton-like daughter is required to satisfy $L_p/(L_p+L_{\pi})> 0.6$.
This selection is about 97\% (95\%) efficient for protons (anti-protons)
while removing about 99\% of pions.

\subsection{B Meson Reconstruction}
Candidate $B$ mesons are reconstructed in the 
$\bp\to\llk$, $\bp\to\llpi$, $\bz\to\llkz$, $\bp \to \llkst$ and $\bz\to\llkstz$ modes.
We use two kinematic variables in the center of mass (CM) frame 
to identify the reconstructed $B$ meson candidates: 
the beam energy constrained mass $\mb = \sqrt{E^2_{\rm beam}-p^2_B}$, 
and the energy difference $\de = E_B - E_{\rm beam}$, 
where $E_{\rm beam}$ is the beam energy, 
and $p_B$ and $E_B$ are the momentum and energy, 
respectively, of the reconstructed $B$ meson.
The candidate region is defined as 
5.2 GeV/c$^2 < \mb < 5.3$ GeV/c$^2$ and $-0.1$ GeV $ < \de< 0.3$ GeV. 
The lower bound in $\de$ for candidate events is chosen to exclude possible background 
from baryonic $B$ decays with higher multiplicities.
From a GEANT~\cite{geant} based Monte Carlo (MC) simulation, 
the signal peaks in a signal box defined by the requirements 
5.27 GeV/c$^2 < \mb < 5.29$ GeV/c$^2$ and $|\de|< 0.05$ GeV.
To ensure {\color{black}that} the decay process be genuinely charmless, we apply a charm veto. 
The regions $2.850$ GeV/c$^2 < M_{\LL} < 3.128$ GeV/c$^2$ 
and $3.315$ GeV/c$^2 < M_{\LL} < 3.735$ GeV/c$^2$ 
are excluded to remove background from modes with $\eta_c, J/\psi$ 
and $\psi^{\prime},\chi_{c0},\chi_{c1}$ mesons, respectively.
According to a study of a rare $B$ decay MC sample, 
the backgrounds in all candidate regions due to self cross-feeds
(e.g between $\bp \to \llk$ and $\bz \to \llkst$) 
or due to other rare decays such as $\bz \to \plpi$, etc., are negligible.  
The contribution of the $B$ background
component with $\Sigma \to \gamma\Lambda$ 
is estimated by fitting the $\de$ distribution.
This will be   
included in the systematic uncertainty from fitting by comparing the results with and 
without this background component in the fit.

\subsection{Background Suppression}
After the above selection requirements,
the background in the fit region arises dominantly from continuum $e^+e^-
\to q\bar{q}$ ($q = u,\ d,\ s,\ c$) processes.
We suppress the jet-like continuum background relative to the more
spherical $B\bar{B}$ signal using a Fisher discriminant~\cite{fisher}.
The Fisher discriminant is a method that combines n-dimensional variables 
into one dimension by weighting linearly; 
the coefficients for each variables are optimized to separate signal and background. 
We optimize the coefficients separately in 7 different missing-mass regions
based on 17 kinematic variables in the CM frame~\cite{ksfw}. The missing-mass
is determined from the rest of the detected particles (treated as
charged pions or photons) in the event assuming they are decay products
of the other $B$ meson. 
{\color{black}
These missing-mass regions are defined as $<$-0.5, -0.5-0.3, 0.3-1.0, 1.0-2.0, 
2.0-3.5, 3.5-6.0, $>$6.0 (GeV/c$^2$).
}
Probability density functions (PDFs) for the Fisher discriminant and
the cosine of the angle between the $B$ flight direction
and the beam direction in the $\Upsilon({\rm 4S})$ rest frame
are combined to form the signal (background)
likelihood ${\mathcal L}_{s}$ (${\mathcal L}_{b}$).
The signal PDFs are determined using signal MC simulation; 
the background PDFs are obtained from the sideband data:
$5.2$ GeV/c$^2$  $ < \mb < 5.26$ GeV/c$^2$ or $0.1 < \de < 0.3$ GeV 
for the $\llk$, $\llpi$, $\llkz$ and $\llkst$ modes;
$5.23$ GeV/c$^2$  $ < \mb < 5.26$ GeV/c$^2$ or $0.1 < \de < 0.2$ GeV 
for the $\llkstz$ mode. 
We require the likelihood ratio 
${\mathcal R} = {\mathcal L}_s/({\mathcal L}_s+{\mathcal L}_b)$ 
to be greater than 0.5, 0.7, 0.3, 0.5 and 0.65 
for the $\llk$, $\llpi$, $\llkz$, $\llkst$ and $\llkstz$ modes, respectively.
These selection criteria are determined by optimization of $n_s/\sqrt{n_s+n_b}$, 
where $n_s$ and $n_b$ denote the expected numbers of signal and background events 
in the signal box, respectively. 
We use the branching fraction $\sim 3 \times 10^{-6}$ ($1 \times 10^{-6}$ for
$\bp \to \llpi$) in the estimation of $n_s$ 
and use the number of data sideband events to estimate $n_b$. 
If there are  multiple $B$ candidates in a single event, 
we select the one with the best ${\mathcal R}$ value.
The fractions of events that have multiple $B$ candidates
are 7.4\%, 12.6\%, 3.7\%, 39.4\% and 26.8\% 
for the $\llk$, $\llpi$, $\llkz$, $\llkst$ and $\llkstz$ modes, respectively.
The systematic errors due to multiple $B$ candidates are described later (Sec.~\ref{MultiCount}).

\section{Extraction of Signal}
\subsection{Unbinned Extended Likelihood Fits}
We perform an unbinned extended likelihood fit 
that maximizes the likelihood function 
\begin{equation}
L = \frac{e^{-(N_{\Lambda\bar{\Lambda}h}+N_{q\bar{q}})}}{N!}
\prod_{i=1}^{N}\left(
N_{\Lambda\bar{\Lambda}h} P_{\Lambda\bar{\Lambda}h}(M_{{\rm bc}_i},\Delta{E}_i)
+N_{q\bar{q}} P_{q\bar{q}}(M_{{\rm bc}_i},\Delta{E}_i) \right).
\end{equation}
to estimate the signal yields for the $\llk$, $\llpi$ and $\llkz$ modes in the candidate region. 
Here $P_{\Lambda\bar{\Lambda}h} (P_{q\bar{q}})$ denotes the signal (background) PDF,
$N$ is the number of events in the fit, and $N_{\Lambda\bar{\Lambda}h}$ and $N_{q\bar{q}}$
are fit parameters representing the number of signal and background yields, respectively. 
The $\llpi$ mode can contain a non-negligible cross-feed contribution from the $\llk$ mode, where the $K^+$ is misidentified as a $\pi^+$.
Hence we include a $\llk$ MC cross-feed shape in the fit for the determination of the $\llpi$ yield. 
The likelihood function {\color{black}is} more complicated for the $\LL K^*$ modes,
\begin{equation}
L = \frac{e^{-(N_{\Lambda\bar{\Lambda}K^*}+N_{\Lambda\bar{\Lambda}K\pi}+N_{q\bar{q}})}}{N!}
\prod_{i=1}^{N}\left(
N_{\Lambda\bar{\Lambda}K^*} P_{\Lambda\bar{\Lambda}K^*}
+N_{\Lambda\bar{\Lambda}K\pi} P_{\Lambda\bar{\Lambda}K\pi}
+N_{q\bar{q}} P_{q\bar{q}}\right).
\end{equation}
since there are contributions from 
non-resonant $B \to \LL K\pi$ decays and  
one more variable in the fit for the $K\pi$ invariant mass, $0.6$ GeV/c$^2 < M_{K\pi} < 1.2$ GeV/c$^2$.

\subsection{Probability Density Functions}
We take each PDF to be the product of shapes in $\mb$ and $\de$ (and $M_{K\pi}$, 
the reconstructed invariant mass of kaon and pion, if applicable), 
which are assumed to be uncorrelated.
Taking $B \to \LL K^*$ for example, for the $i$-th event, $P_{\LL K^*} = 
P_{\mb}(M_{{\rm bc}_i}) \times P_{\de}(\Delta{E}_i) 
\times P_{K\pi}(M_{{K\pi}_i})$.
For the PDFs related to $B$ decays, 
we use a Gaussian function to represent $P_{\mb}$ and a double Gaussian for $P_{\de}$ 
with parameters determined from MC signal simulation.
{\color{black}
The theoretical p-wave Breit-Wigner resonance
}
function is defined by 
Eqns.~\ref{eq:pwave-1}, \ref{eq:pwave-2} and \ref{eq:pwave-3}, 
where A is a normalization factor, 
$\Gamma$ is the width of the peak and 
$m_{K^*}, m_K$ and $m_{\pi}$ are the nominal masses of the 
$K^*, K$ and $\pi$~\cite{PDG}, respectively.

\begin{equation}
BW({\rm p-wave})=A \times \frac{m_{K^*} \times \Gamma \times (\frac{q}{q_0})^3}
           {(M_{K\pi}^2-m_{K^*}^2)^2 + (m_{K^*} \times \Gamma \times (\frac{q}{q_0})^3)^2},
\label{eq:pwave-1}
\end{equation}

\begin{equation}
{\rm where}~~~q = \sqrt{\left(\frac{M_{K\pi}^2 + m_{\pi}^2 - m_K^2}{2M_{K\pi}}\right)^2-m_{\pi}^2},
\label{eq:pwave-2}
\end{equation}

\begin{equation}
~~~~~~~~~~q_0 = \sqrt{\left(\frac{m_{K^*}^2 + m_{\pi}^2 - m_K^2}{2m_{K^*}}\right)^2-m_{\pi}^2}.
\label{eq:pwave-3}
\end{equation}
We use these functions to parameterize the $P_{M_{K\pi}}$ 
distributions for $\kst$ and $\kstz$, 
and use a LASS function obtained from the LASS collaboration~\cite{LASS}
to model the nonresonant $P_{M_{K\pi}}$ distribution.
For the continuum background PDFs, 
we use a parameterization that was first used by the ARGUS collaboration~\cite{Argus}, 
$ f(\mb)\propto x\sqrt{1-x^2}e^{-\xi (1-x^2)}$,  
to model the $P_{\mb}$ distribution with $x$ given by $\mb/E_{\rm beam}$ 
and where $\xi$ is a fit parameter. 
The $P_{\de}$ distribution is modeled by a normalized second order polynomial 
whose coefficients are fit parameters. The continuum background PDF for the $K^*$ modes,
$P_{M_{K\pi}}$, is modeled by a p-wave function and a threshold function, 
$P_{M_{K\pi}} = r\times P_{p-\textrm{wave}} + (1-r)\times P_{\textrm{threshold}}$ and
$P_{\textrm{threshold}} \propto (M_{K\pi}-m_{K}-m_{\pi})^{s}\times
e^{[c_1\times(M_{K\pi}-m_{K}-m_{\pi})+c_2\times(M_{K\pi}-m_{K}-m_{\pi})^2]}$
where $r, s, c_1$ and $c_2$ are fit parameters.

\vskip 0.5cm
\begin{figure}[p]
 {\hskip -13.5cm {\bf{(a)}}}
 {\vskip  -1.5cm {\includegraphics[width=0.78\textwidth]{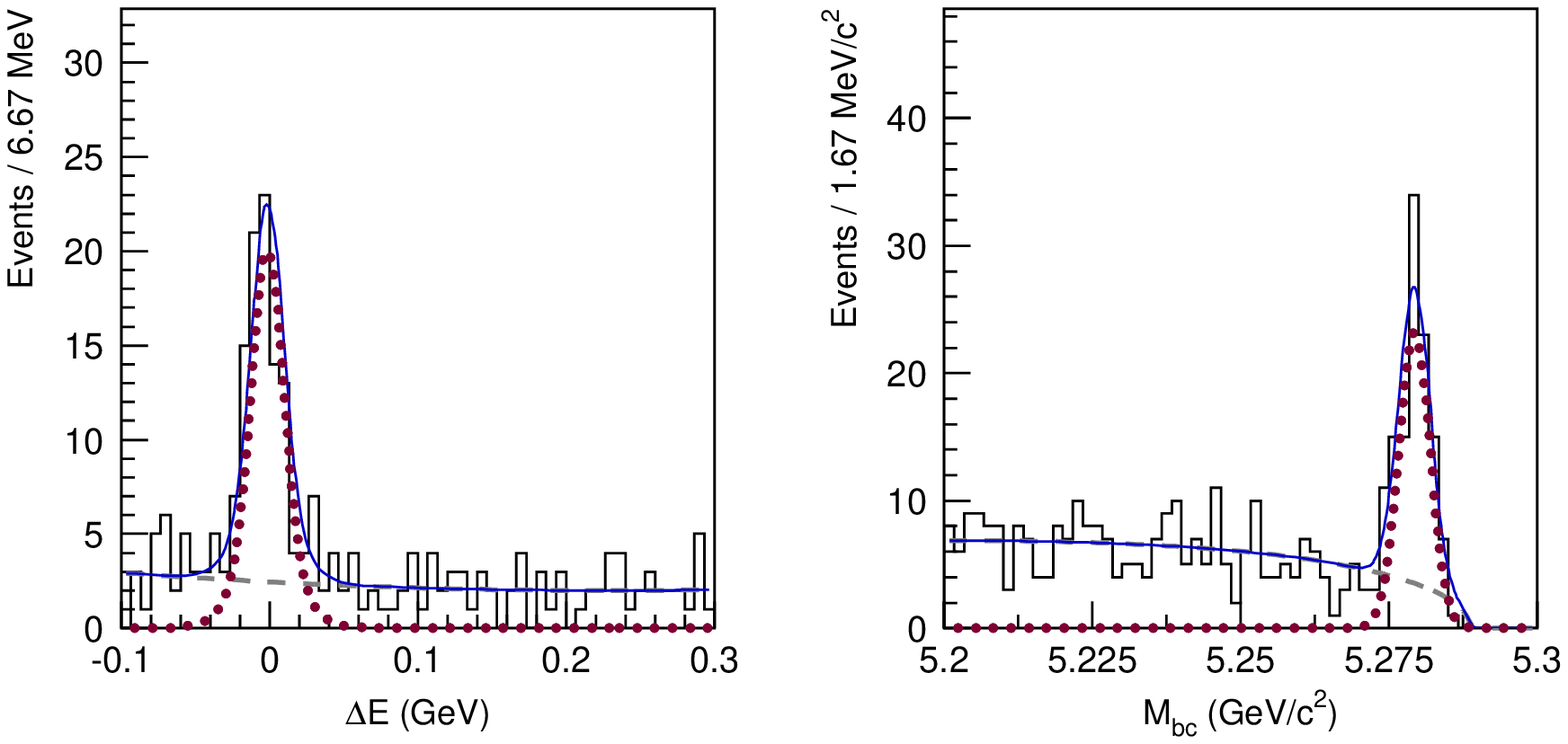}}}\\
 {\hskip -13.5cm {\bf{(b)}}}
 {\vskip  -1.5cm {\includegraphics[width=0.78\textwidth]{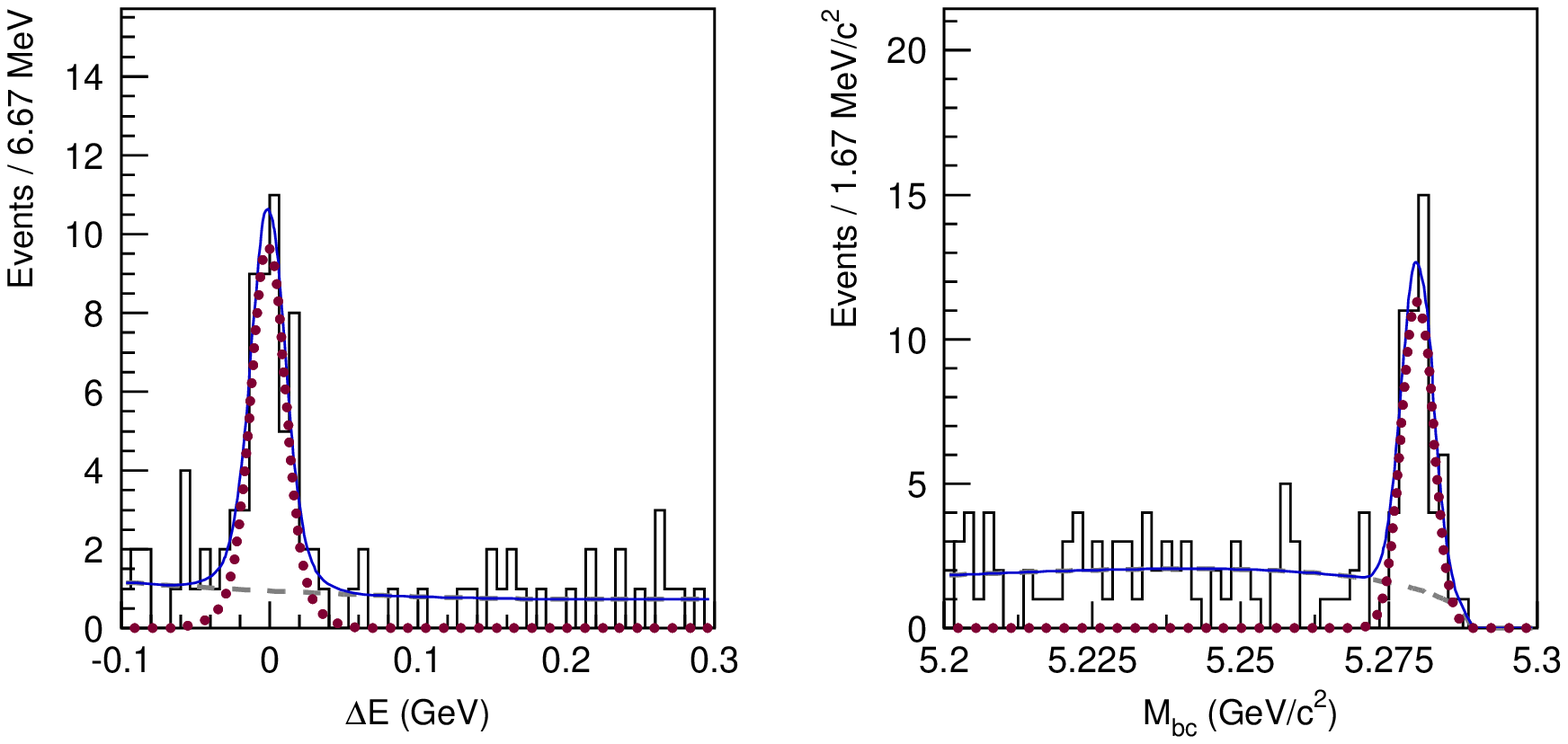}}}\\
 {\hskip -13.5cm {\bf{(c)}}}
 {\vskip  -1.5cm {\includegraphics[width=0.78\textwidth]{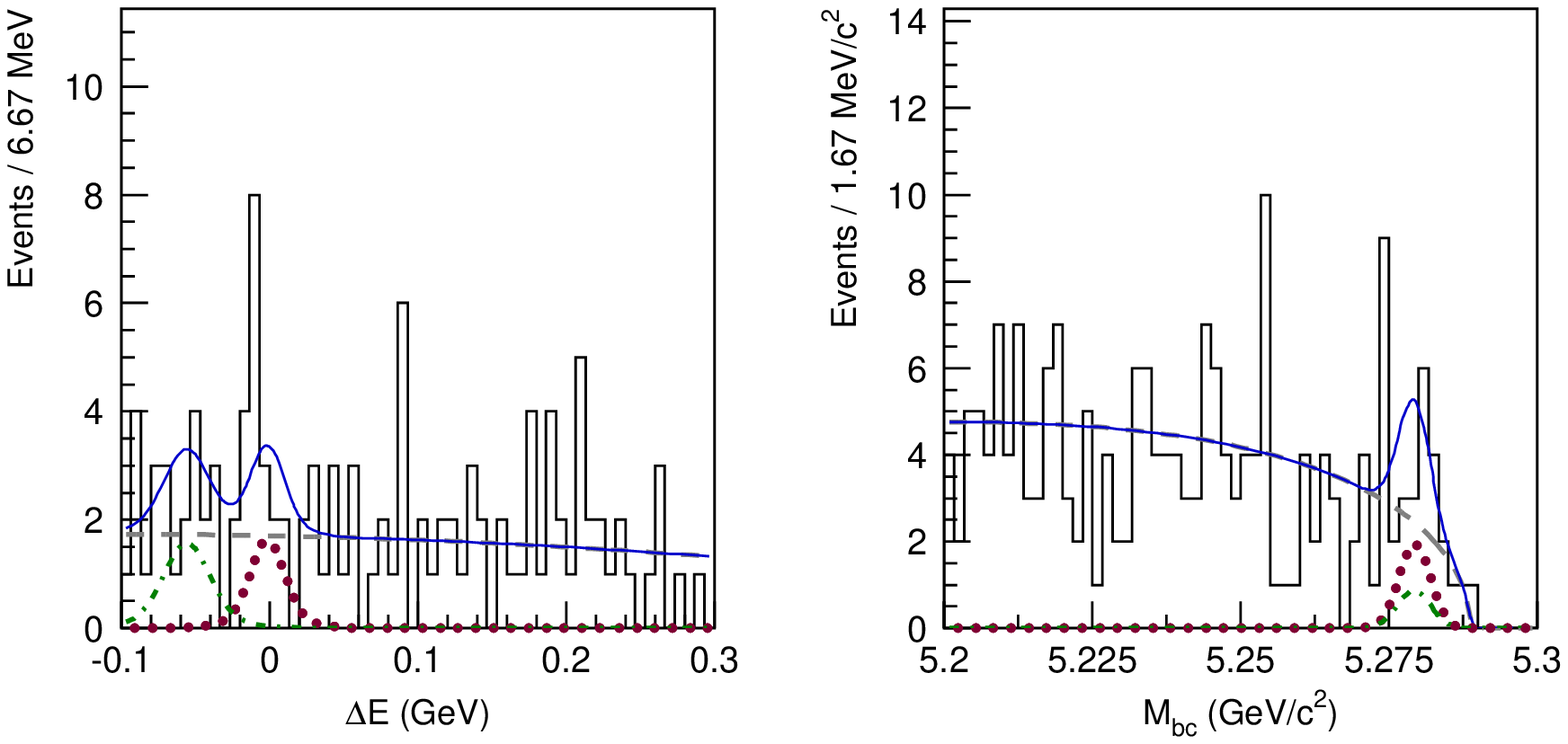}}}
 \caption{Distributions of
 $\de$ (with $5.27$ GeV/c$^2 < \mb < 5.29$ GeV/c$^2$) and
 $\mb$ (with $|\de| < 0.05$ GeV) 
 for (a) $\bp \to \llk$, (b) $\bz \to \llkz$ and (c) $\bp \to \llpi$ modes.
 The dibaryon mass $\mll$ is required to be less than 2.85 GeV/c$^2$.
 The solid curves, dotted curves, and dashed curves represent the total fit
 result, fitted signal and fitted background, respectively.
 The dot-dashed curves in plot (c) show the background contribution from the $\bp \to \llk$ mode.}
 \label{fg:mergembde-1}
\end{figure}

\vskip 1.5cm
\begin{figure}[htp]
 {\hskip -16cm {\bf{(a)}}}
 {\vskip -1cm {\includegraphics[width=1.0\textwidth]{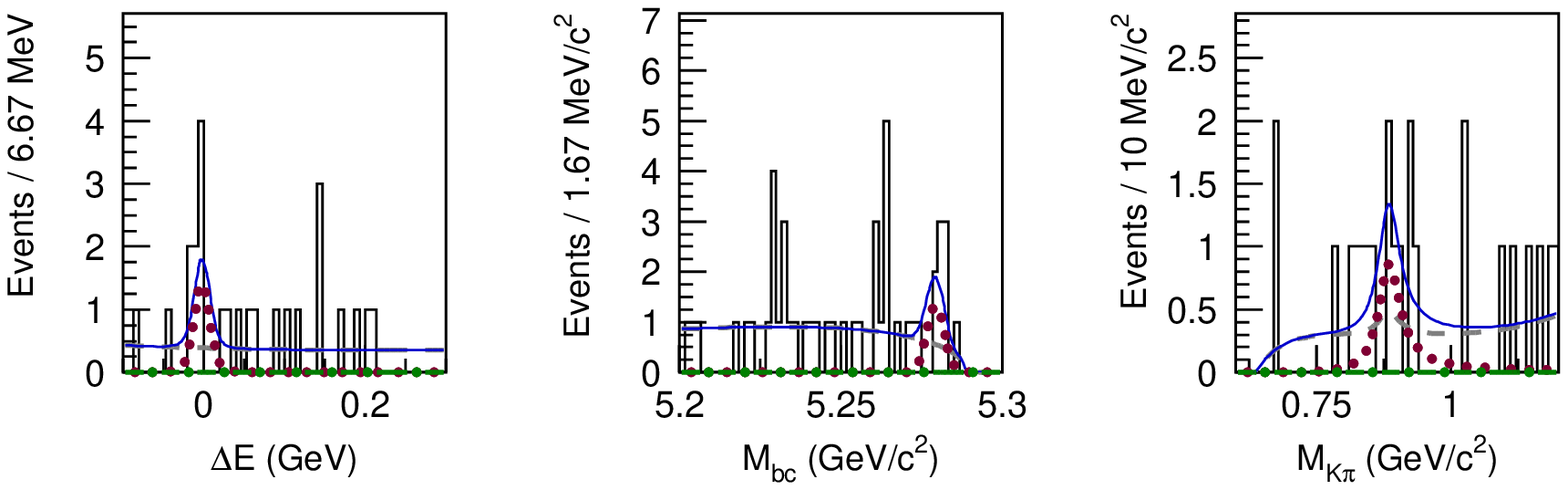}}}\\
 {\vskip -1.5cm {\hskip -16cm {\bf{(b)}}}}
 {\vskip -1cm {\includegraphics[width=1.0\textwidth]{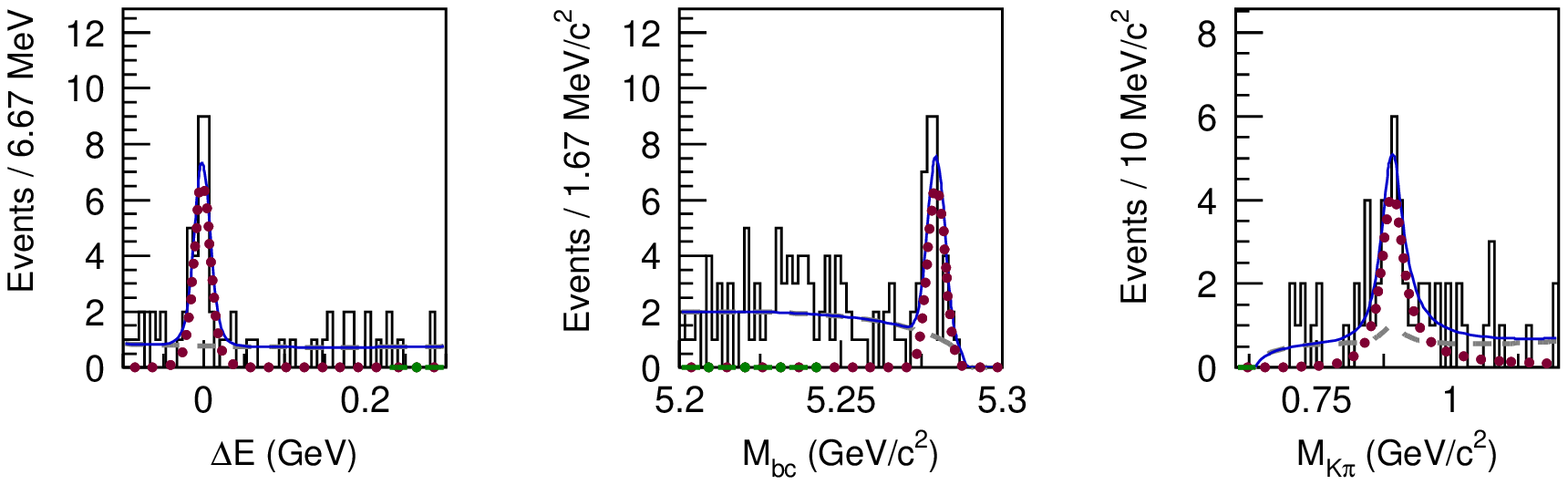}}}
 \caption{Distributions of
 $\de$ (with $5.27$ GeV/c$^2 < \mb < 5.29$ GeV/c$^2$ and $0.816 < M_{K\pi} < 0.976$ GeV/c$^2$),
 $\mb$ (with $|\de| < 0.05$ GeV and $0.816$ GeV/c$^2 < M_{K\pi} < 0.976$ GeV/c$^2$) 
 and $M_{K\pi}$ (with $|\de| < 0.05$ GeV and $5.27$ GeV/c$^2 < \mb < 5.29$ GeV/c$^2$)
 for (a) $\bp \to \llkst$ and (b) $\bz \to \llkstz$ modes in the threshold-mass-enhanced region. 
 The solid curves, dotted curves, and dashed curves represent the total fit
 result, fitted signal and fitted background, respectively.}
 \label{fg:mergembde-2}
\end{figure}

\section{Physics Results}
\subsection{Fitting Results}
Figures~\ref{fg:mergembde-1} and~\ref{fg:mergembde-2} show the fit results
for $\bp \to \llk$, $\bz \to \llkz$, 
$\bp \to \llpi$, 
$\bp \to \llkst$ and $\bz \to \llkstz$
in the $\mll$ region below 2.85 GeV/c$^2$,
which we refer to as the threshold-mass-enhanced region.
The resulting signal yields are given in Table~\ref{number-of-yield}.
The significance is defined as $\sqrt{-2{\rm ln}(L_0/L_{\rm max})}$,
where $L_0$ and $L_{\rm max}$ are the likelihood values returned by the 
fit with the signal yield fixed to zero and at its best fit value.
These values include the systematic uncertainty obtained by 
varying signal PDF parameters by their 1 $\sigma$ errors.

\begin{table}[htb]
  \vskip 1.0cm
 \caption{Signal yields for each decay mode with $\mll <$ 2.85 GeV/c$^2$.}
 \begin{center}
  \begin{tabular}{|c|ccccc|}
   \hline
   Mode & $\llk$ & $\llkz$ & $\llpi$ & $\llkst$ & $\llkstz$\\
   \hline
   Yield & $92.7^{+11.0}_{-10.3}$ & $45.8^{+7.6}_{-6.9}$ & $7.76^{+4.49}_{-3.72}$ & $6.54^{+3.37}_{-2.63}$ & $31.4^{+7.4}_{-6.8}$\\
   Significances~($\sigma$) & 16.4 & 12.4 & 2.5 & 3.7 & 9.3\\
   \hline
  \end{tabular}
 \end{center}
 \label{number-of-yield}
\end{table}

\begin{figure}[pht]
 \vskip 0.10cm
 \begin{center}
 \hskip -6.cm {\bf (a)} \hskip 6cm {\bf (b)}\hskip 4.5cm\\
 \vskip -0.6cm
 \includegraphics[width=0.43\textwidth]{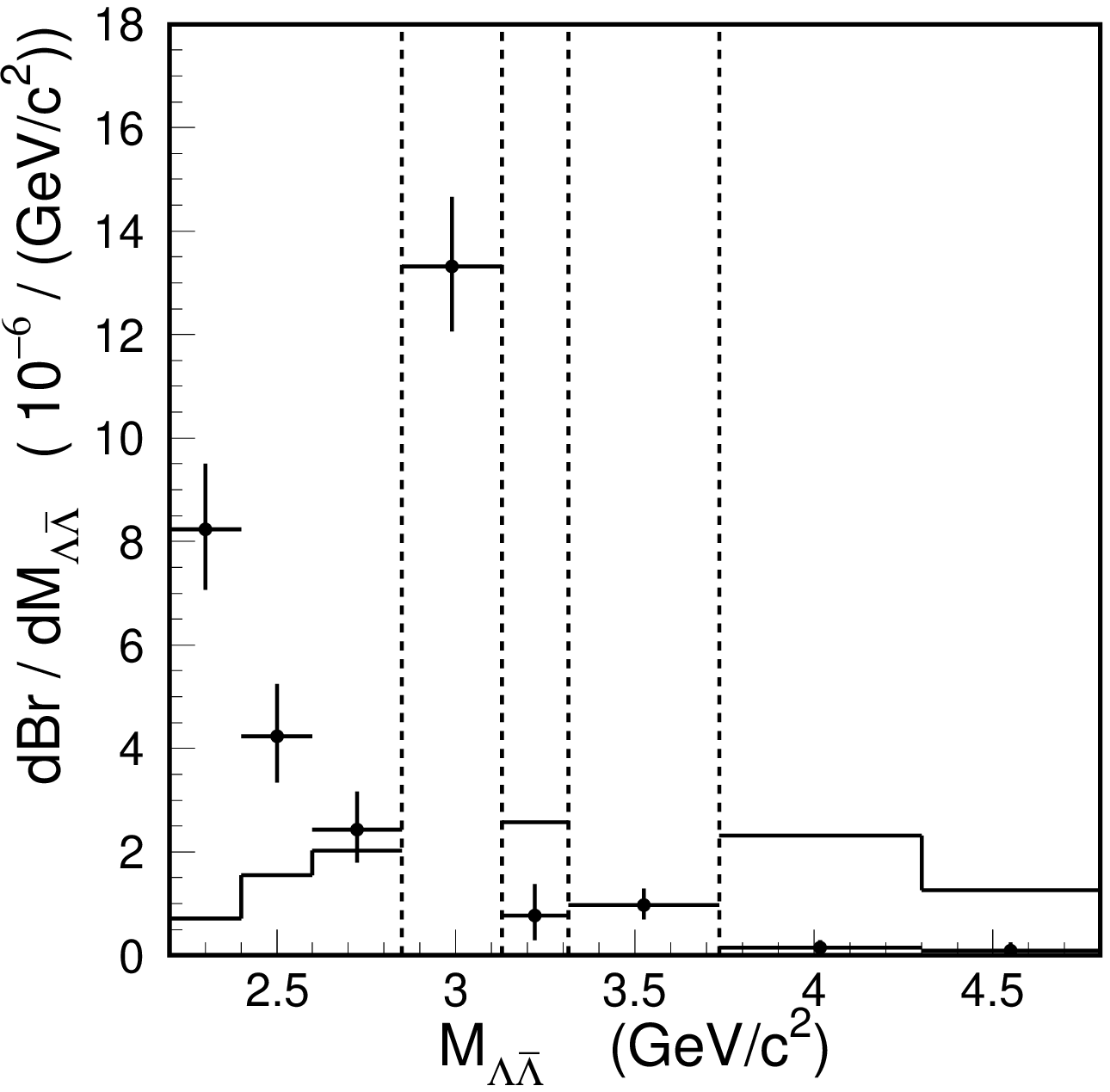}
 \includegraphics[width=0.43\textwidth]{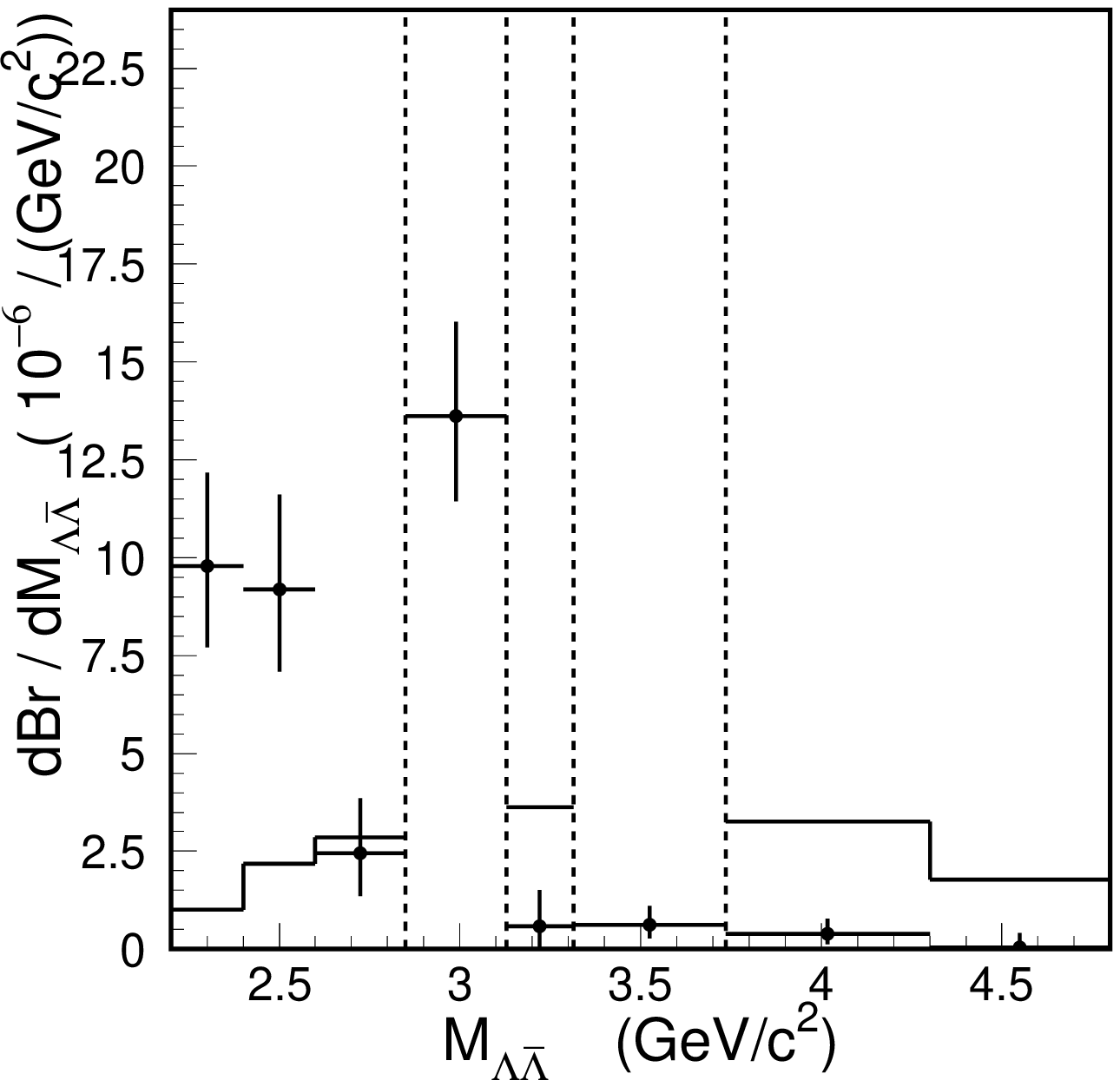}\\
 \hskip -7 cm
 {\bf{(c)}}
 \vskip -0.5cm
 \includegraphics[width=0.43\textwidth]{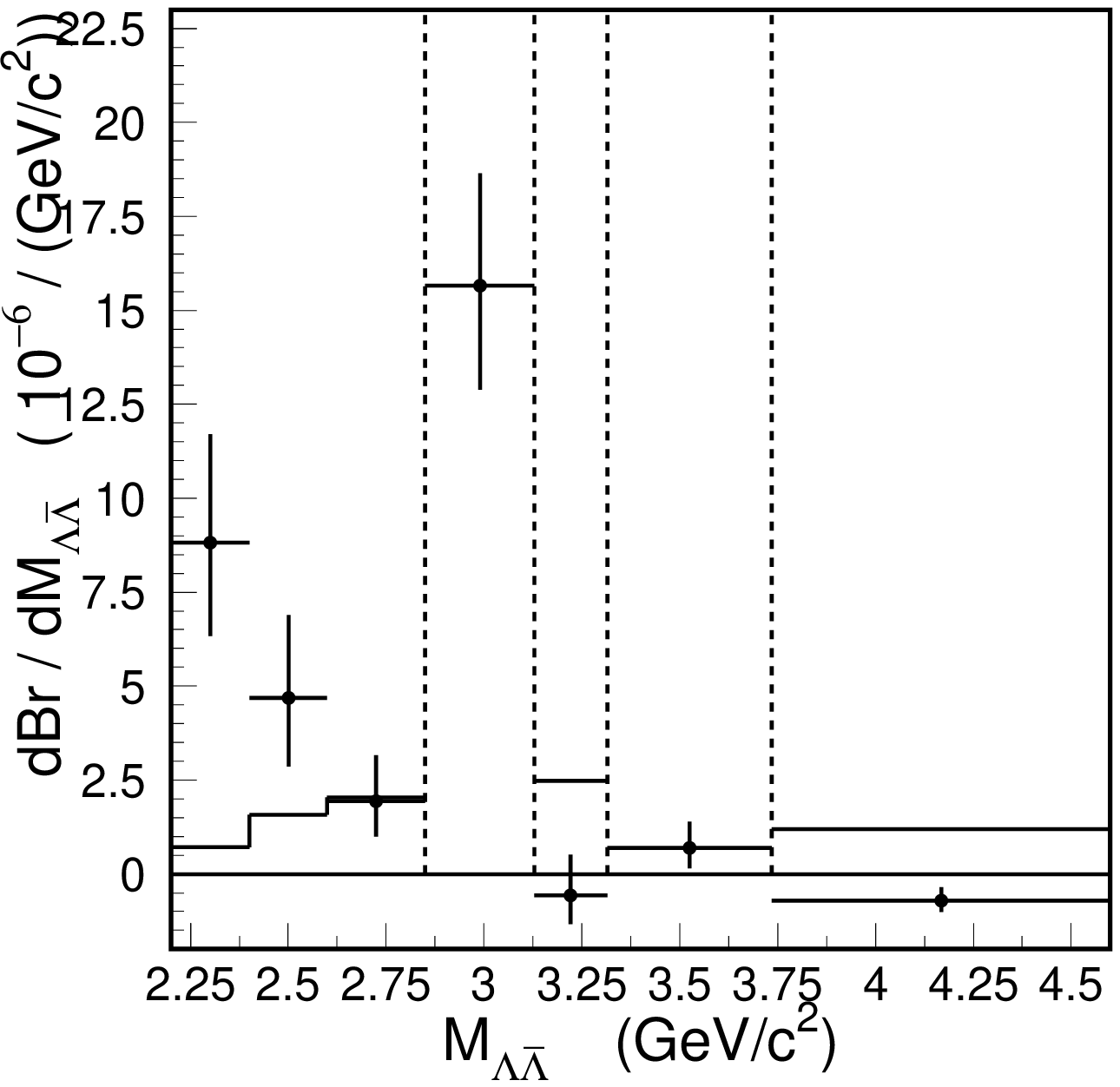}
 \caption{Differential branching fractions for
 (a) $\bp \to \llk$, (b) $\bz \to \llkz$ and (c) $\bz \to \llkstz$ modes 
 as a function of $\mll$.
 Note that two bins with
 2.850 GeV/c$^2$ $< \mll <$ 3.128 GeV/c$^2$ and 3.315 GeV/c$^2$ $< \mll <$ 3.735 GeV/c$^2$
 contain charmonium events and 
 are excluded from the charmless signal yields.
 The solid histograms are from phase space MC simulation 
 with area normalized to the charmless signal yield.
 }
 \label{fg:allphase}
 \end{center}
\end{figure}

\begin{table}[p]
 \caption{Signal yields and branching fractions ${\cal B}$ ($10^{-6}$) 
 in different $\mll$ regions for $\bp\to\llk$ and $\bz\to\llkz$ decays.
 The $\dag$ symbol indicates a charm veto bin.}
 \begin{center}
  \begin{tabular}{|l|cc|cc|}
  \hline
                       & \multicolumn{2}{c|}{$\bp \to \llk$}              & \multicolumn{2}{c|}{$\bz \to \llkz$} \\
   $\mll$ (GeV/c$^2$)  & Yield                   & ${\cal B}$ ($10^{-6}$) & Yield                   & ${\cal B}$ ($10^{-6}$)\\
   \hline
   $<2.4$              & $50.6^{+7.9}_{-7.2}$    & $1.65^{+0.26}_{-0.23}$ & $20.8^{+5.1}_{-4.4}$    & $1.96^{+0.48}_{-0.41}$\\
   $2.4-2.6$           & $24.7^{+5.9}_{-5.2}$    & $0.85^{+0.20}_{-0.18}$ & $18.3^{+4.8}_{-4.2}$    & $1.84^{+0.49}_{-0.42}$\\
   $2.6-2.85$          & $17.5^{+5.3}_{-4.6}$    & $0.61^{+0.18}_{-0.16}$ & $ 5.9^{+3.4}_{-2.7}$    & $0.61^{+0.35}_{-0.28}$\\
   $2.85-3.128$~($^{\dag}$)    & $117.5^{+11.8}_{-11.2}$ & $3.70^{+0.37}_{-0.35}$ & $39.3^{+7.0}_{-6.3}$    & $3.79^{+0.67}_{-0.60}$\\
   $3.128-3.315$       & $ 5.0^{+3.9}_{-3.1}$    & $0.14^{+0.11}_{-0.09}$ & $ 1.2^{+2.0}_{-1.2}$    & $0.11^{+0.17}_{-0.10}$\\
   $3.315-3.735$~($^{\dag}$)   & $16.0^{+5.2}_{-4.5}$    & $0.41^{+0.13}_{-0.11}$ & $ 3.0^{+2.4}_{-1.7}$    & $0.26^{+0.21}_{-0.15}$\\
   $3.735-4.3$         & $ 3.7^{+3.8}_{-3.0}$    & $0.08^{+0.09}_{-0.07}$ & $ 2.6^{+2.5}_{-1.8}$    & $0.22^{+0.21}_{-0.15}$\\
   $>4.3$              & $ 1.9^{+3.3}_{-2.5}$    & $0.05^{+0.08}_{-0.06}$ & $ 0.3^{+1.8}_{-1.0}$    & $0.03^{+0.18}_{-0.11}$\\
   \hline 
   charmless           & $103.4^{+12.9}_{-11.2}$ & $3.38^{+0.41}_{-0.36}$ & $49.1^{+8.6}_{-7.0}$    & $4.76^{+0.84}_{-0.68}$\\
   \hline
  \end{tabular}
 \end{center}
 \label{mll-bins-k}
\end{table}

\begin{table}[p]
 \caption{Signal yields and branching fractions ${\cal B}$ ($10^{-6}$) 
 in different $\mll$ regions for $\bz\to\llkstz$ decay.
 The $\dag$ symbol indicates a charm veto bin.} 
 \begin{center}
  \begin{tabular}{|l|cc|}
   \hline
                       & \multicolumn{2}{c|}{$\bz \to \llkstz$}\\
   $\mll$ (GeV/c$^2$)  & Yield                & ${\cal B}$ ($10^{-6}$)\\
   \hline 
   $<2.4$              & $19.6^{+6.4}_{-5.5}$ & $1.76^{+0.58}_{-0.50}$\\
   $2.4-2.6$           & $ 9.6^{+4.5}_{-3.7}$ & $0.94^{+0.44}_{-0.37}$\\
   $2.6-2.85$          & $ 4.6^{+2.9}_{-2.2}$ & $0.48^{+0.31}_{-0.23}$\\
   $2.85-3.128$~($^{\dag}$)  & $43.8^{+8.3}_{-7.7}$ & $4.35^{+0.83}_{-0.77}$\\
   $3.128-3.315$       & $-1.2^{+2.4}_{-1.7}$ & $-0.10^{+0.20}_{-0.14}$\\
   $3.315-3.735$~($^{\dag}$) & $ 3.5^{+3.6}_{-2.7}$ & $0.29^{+0.29}_{-0.22}$\\
   $>3.735$            & $-7.2^{+3.6}_{-3.1}$ & $-0.61^{+0.31}_{-0.26}$\\
   \hline
   charmless           & $25.3^{+9.4}_{-7.8}$ & $2.46^{+0.87}_{-0.72}$\\
   \hline
  \end{tabular}
 \end{center}
 \label{mll-bins-kstz}
\end{table}

\subsection{Observed Branching Fractions}
\subsubsection{Branching Fractions}
\begin{figure}[hbt]
\begin{center}
\includegraphics[width=0.43\textwidth]{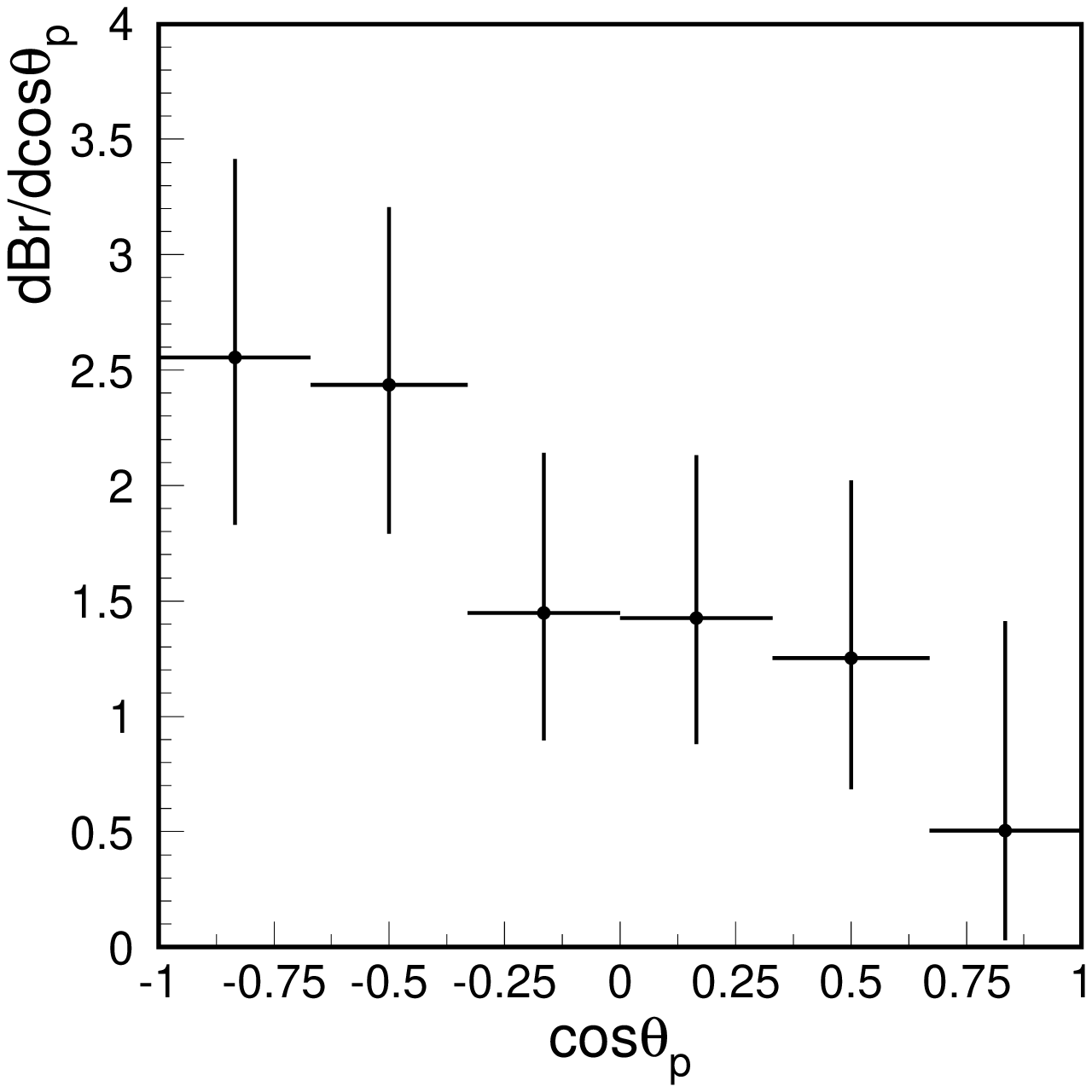}
\vskip -0.8cm
\caption{
Differential branching fractions {\it vs.} $\cos\theta_{p}$ for  $B^+ \to \LL K^+$ in the threshold-mass-enhanced region.}
\label{fg:llk-costhetap}
\end{center}
\end{figure}

The differential branching fractions as a function of $\mll$ 
for the observed modes are shown in Fig.~\ref{fg:allphase}. 
Tables~\ref{mll-bins-k} and \ref{mll-bins-kstz} give the yields 
and the corresponding branching fractions for each $\mll$ bin.
The yields are obtained
from ($\de$, $\mb$(, $M_{K\pi}$)) unbinned extended maximum likelihood fits 
for each bin of $\mll$.
We find that a threshold enhancement
is also present for $\bz \to \llkstz$ and $\bz \to \llkz$ decays. 
We sum the charmless partial branching fractions, 
where the summation excludes bins in the two charmonium regions, 
to obtain:
${\mathcal B}(\bp \to \llk) = (3.38^{+0.41}_{-0.36} \pm 0.41) \times 10^{-6}$, 
${\mathcal B}(\bz \to \llkz) = (4.76^{+0.84}_{-0.68} \pm 0.61) \times 10^{-6}$, and 
${\mathcal B}(\bz \to \llkstz) = (2.46^{+0.87}_{-0.72} \pm 0.34 )\times 10^{-6}$. 
For the $\llkst$ mode, 
we find $\mathcal{B} (\bp \to \llkst) = (2.19^{+1.13}_{-0.88} 
\pm 0.33)  \times 10^{-6}$ with 3.7$\sigma$ significance in the threshold-mass-enhanced region
using the yield in Table~\ref{number-of-yield}.
The differential branching fractions are obtained by correcting the yields 
for the $\mll$ dependent efficiency, which is estimated from signal MC.
\label{sec:llk-costhetap}
{\color{black}
Here, we include the efficiency correction for $\Lambda$ polarization
reported in Ref.~\cite{Wang, Suzuki}
as our default MC does not include such an effect.
}
The correction factors are 1.17, 1.23, 1.20, 1.22, and 1.16
for $\bp\to\llk$, $\bp\to\llpi$, $\bz\to\LL\kz$,
$\bz\to\llkstz$ and $\bp\to\llkst$, respectively.
{\color{black} These factors are obtained in a model independent way.
We first use the phase space MC sample to obtain the efficiency function in 
$\cos\theta_p$, where $\cos\theta_p$ is the polar angle of proton in the $\Lambda$ 
helicity frame.
We then use the 
$\cos\theta_p$ distributions in the data sideband and signal regions to find
their corresponding average efficiencies. With the signal yield information from the 
fit, the model independent signal efficiency can be estimated. 
Fig.~\ref{fg:llk-costhetap} shows the differential branching fractions in bins
of $\cos\theta_p$ for $B^+ \to \LL K^+$. This distribution is not flat but 
does agree with the theoretical expectation~\cite{Suzuki}.}    

To verify the branching fraction measurement procedure, 
we use $J/\psi$ $K^{(*)}$ events with 
$J/\psi \to \LL$ in the region $3.070~{\rm GeV/c^2}< \mll < 3.125~{\rm GeV/c^2}$.
{\color{black} Using $\mathcal B(J/\psi \to \LL) = ( 1.61 \pm 0.15 ) \times 10^{-3}$~\cite{PDG}, }
we obtain branching fractions 
{\color{black}
of 
$(1.30^{+0.21}_{-0.20}) \times 10^{-3}$, 
$(0.66^{+0.23}_{-0.19}) \times 10^{-3}$, and 
$(2.08^{+0.45}_{-0.42}) \times 10^{-3}$} for 
$\bp \to J/\psi K^+$, $\bz \to J/\psi K^0$, $\bz \to J/\psi K^{*0}$, respectively, 
which agree with the world average values~\cite{PDG}
within $2 \sigma$ including systematic errors that are similar to those for the signal mode discussed below.
%

\subsubsection{Polar Angle Distribution }
\label{sec:PolarAngleD}
\begin{figure}[bht]
\vskip 0.5cm
\begin{center}
\vskip -1.2cm
\includegraphics[width=0.42\textwidth]{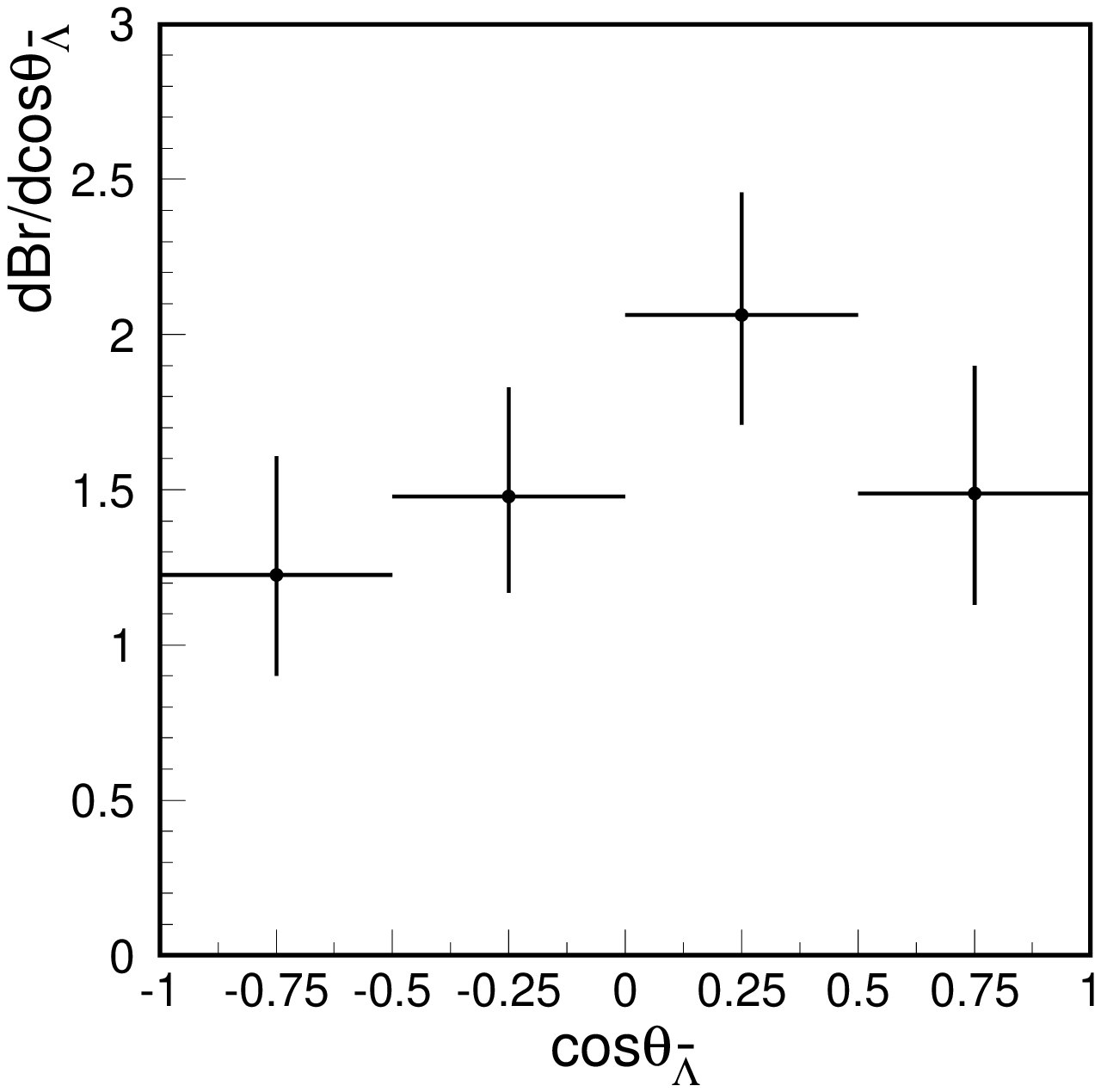}
\caption{
Differential branching fractions
{\it vs.} $\cos \theta_{\bar{\Lambda}}$
 in the $\LL$ pair system for $\bp \to \llk$ in the threshold-mass-enhanced region.
}
\label{fg:llk-costhl}
\end{center}
\end{figure}

  Figure~\ref{fg:llk-costhl} shows the angular distribution of the $\bar{\Lambda}$ in the $\LL$ rest frame
  for the threshold-mass-enhanced region. 
The {\color{black}yields are} obtained
from ($\de$, $\mb$) unbinned extended maximum likelihood fits 
for each bin of $\cos\theta_{\bar{\Lambda}}$.
  The angle $\theta_{\bar{\Lambda}}$ is defined as the angle between 
  the $\bar{\Lambda}$ direction and the $K^+$ direction in the $\LL$ pair rest frame.  
Here, we make a $\cos\theta_{\bar{\Lambda}}$ dependent efficiency correction 
and an average correction for the $\bar{\Lambda}$ helicity dependence 
as discussed above.
  The distribution shows no significant forward peak, 
  in contrast to the prominent peak reported in $\bp \to \ppk$~\cite{polar}, 
  which is a unique signature of the intriguing result discussed above.

\subsubsection{Helicity Distribution}
\begin{figure}[htb]
\begin{center}
\vskip -0.7cm
\includegraphics[width=0.43\textwidth]{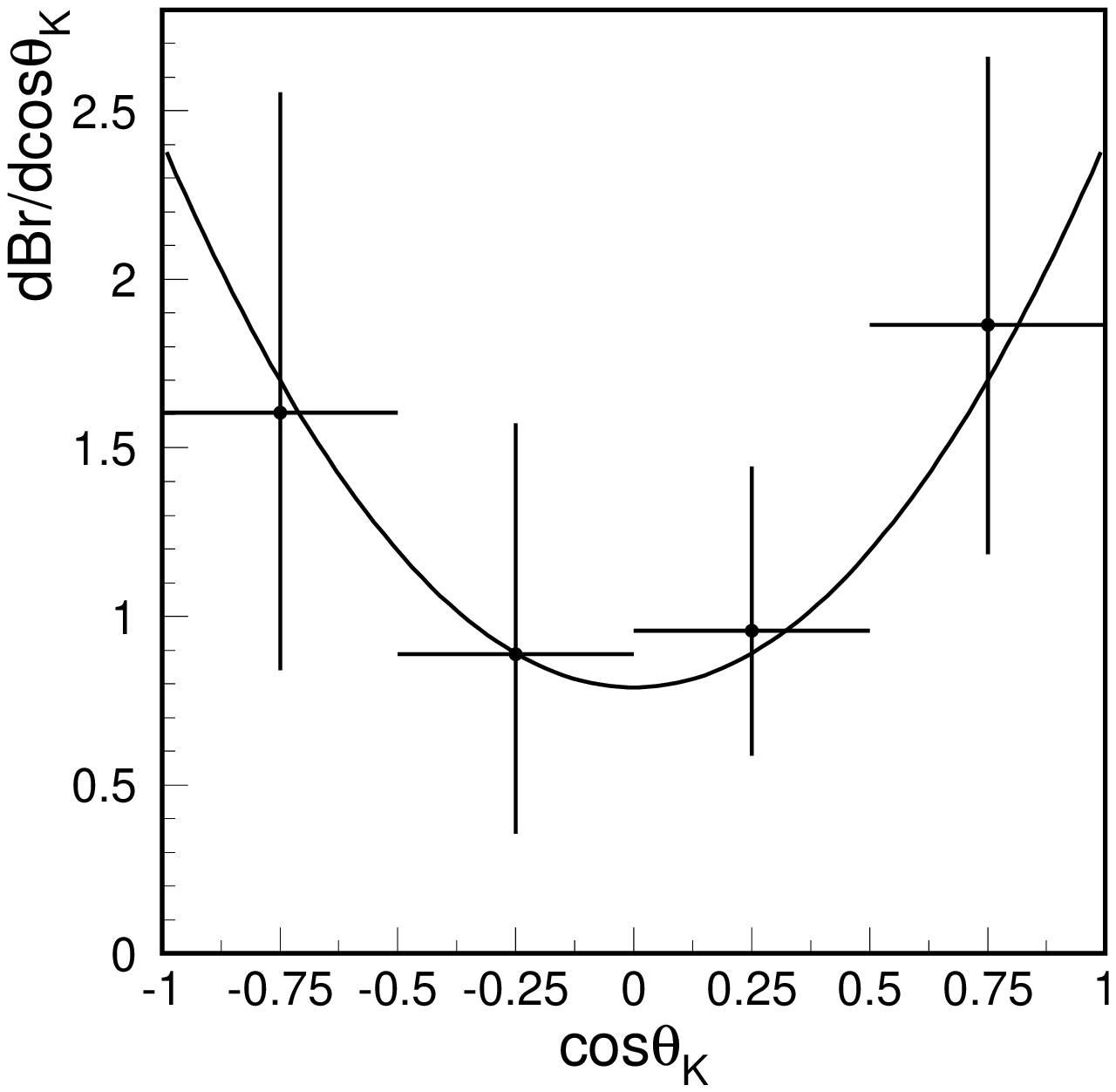}
\caption{
Differential branching fractions
{\it vs.} $\cos \theta_K$
 in the $K^{*0}$  system for $\bz \to \llkstz$ in the threshold-mass-enhanced region. 
 The solid curve is the result of the fit.
}
\label{fg:llk-helicity}
\end{center}
\end{figure}

We study the $K^{*0}$ polarization in $\llkstz$ decay,
as the $\kstz$ meson is found to be
almost 100\% polarized with a fraction of $(101 \pm 13 \pm 3)\%$ in the
helicity zero state in $\bz \to \ppkstz$ decay~\cite{JHChen}. 
To study the $K^{*0}$ polarization, 
we use {\color{black}a} MC simulation to obtain the efficiency as a function of 
$\cos\theta_{K}$ in the threshold-mass-enhanced region, 
where the angle $\theta_{K}$ is defined as the angle between 
the opposite $B$ direction and the $K^+$ direction in the $K^{*0}$ rest frame.
We separate the $\cos\theta_{K}$ distribution into 4 bins for data.
We then use ($\de$, $\mb$, $M_{K\pi}$) unbinned extended maximum likelihood fits 
to obtain signal yields in bins of $\cos\theta_K$ 
and calculate the branching fractions for each bin with the corresponding efficiency.
Finally, we use $D^1_{00}$ and $D^1_{10}$ functions, i.e. 
$\frac{3}{2}\cos^2\theta_K$ for a pure helicity zero state and 
$\frac{3}{4}\sin^2\theta_K$ for a pure helicity one ($\pm 1$) state, 
to fit this branching fraction distribution. 
The fit result is shown in Fig.~\ref{fg:llk-helicity}. 
We find that the $\kstz$ meson 
{\color{black}
is polarized with 
}
$(60 \pm 22 \pm 8)\%$ in the helicity zero state. 

\subsubsection{Upper Limits and Interpretation}
{\color{black}
  For modes with signal significance less than 4$\sigma$, we set the corresponding 
  upper limits on the decay branching fractions at the 90\% confidence level in the 
  threshold-mass-enhanced region. 
  Using the methods described in Refs.~\cite{Gary, Conrad}, 
  we obtain ${\mathcal B}(\bp \to \llkst) < 4.98 \times 10^{-6}$,
  where the systematic uncertainty has been taken into account.


  Naively, one would expect that the 
  ratio of $\mathcal{B}(\bp \to \llpi)$ to $\mathcal{B}(\bp \to \llk)$ is similar to
  the one of $\mathcal{B}(\bp \to \pppi)$ to $\mathcal{B}(\bp \to \ppk)$~\cite{Wei}.
  The anticipated signal yield for $ \bp \to \llpi$ is $26.83^{+4.46}_{-4.13}$.
  However, we find no significant signal in the 
  threshold-mass-enhanced region for $\bp \to \llpi$ and obtain the upper limit
  ${\mathcal B}(\bp \to \llpi) < 0.94 \times 10^{-6}$ at the 90\% confidence level.
  As a cross-check,
  we measure $\mathcal{B} (\bp \to \llk) = (3.57^{+1.82}_{-1.54}) \times 10^{-6}$ 
  using the misidentified $\bp \to \llk$ component in the fit. 
  This value agrees well with our ${\mathcal B}(\bp \to \llk)$ measurement.}

\begin{figure}[htp]
\begin{center}
\vskip -0.6cm
\includegraphics[width=0.75\textwidth]{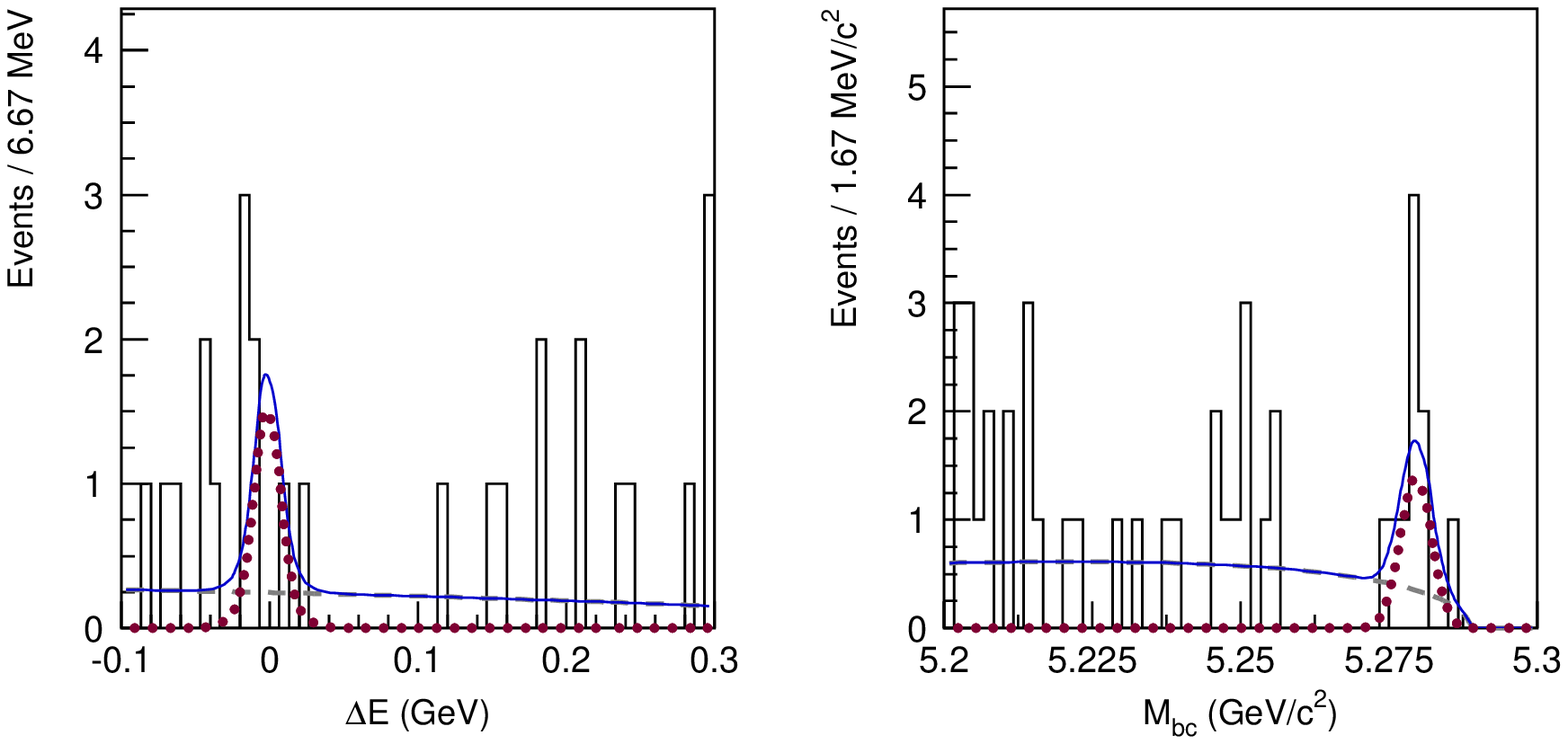}
\vskip -0.6cm
\caption{
Distributions of
$\de$ (with $5.27$ GeV/c$^2 < \mb < 5.29$ GeV/c$^2$) and
$\mb$ (with $|\de| < 0.05$ GeV) 
for the $\bz \to \LL \bar{D^0}$ mode.
The result includes the whole $\mll$ region.
The solid curves, dotted curves, and dashed curves show the total fit 
result, fitted signal and fitted background, respectively.
}
\label{fg:lld0-fitting}
\end{center}
\end{figure}

\begin{figure}[htp]
\vskip -0.3cm
\begin{center}
\hskip -7.4cm {\bf (a)} \hskip 6.6cm {\bf (b)}\\
\vskip -1.8cm
\includegraphics[width=0.41\textwidth]{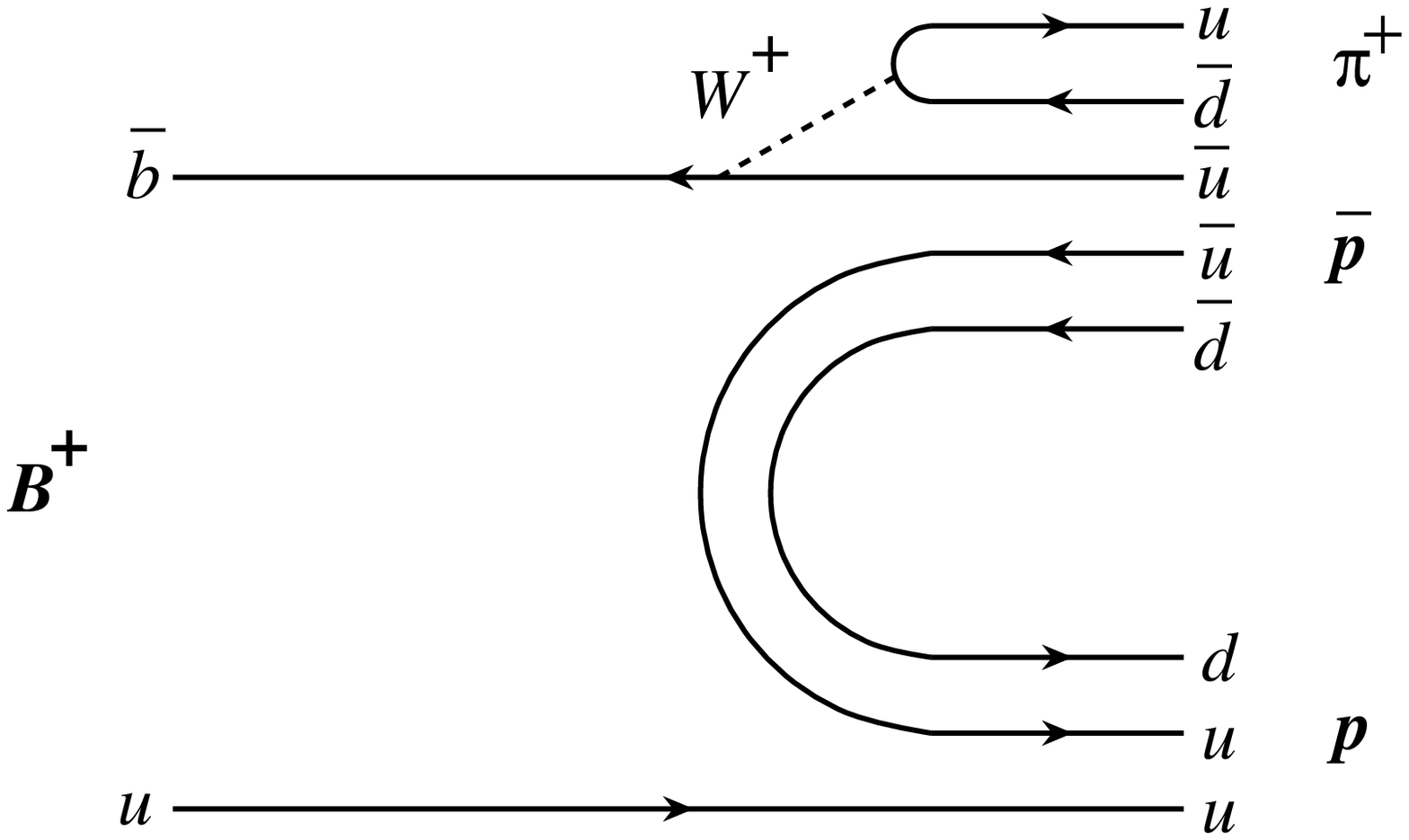}
\includegraphics[width=0.41\textwidth]{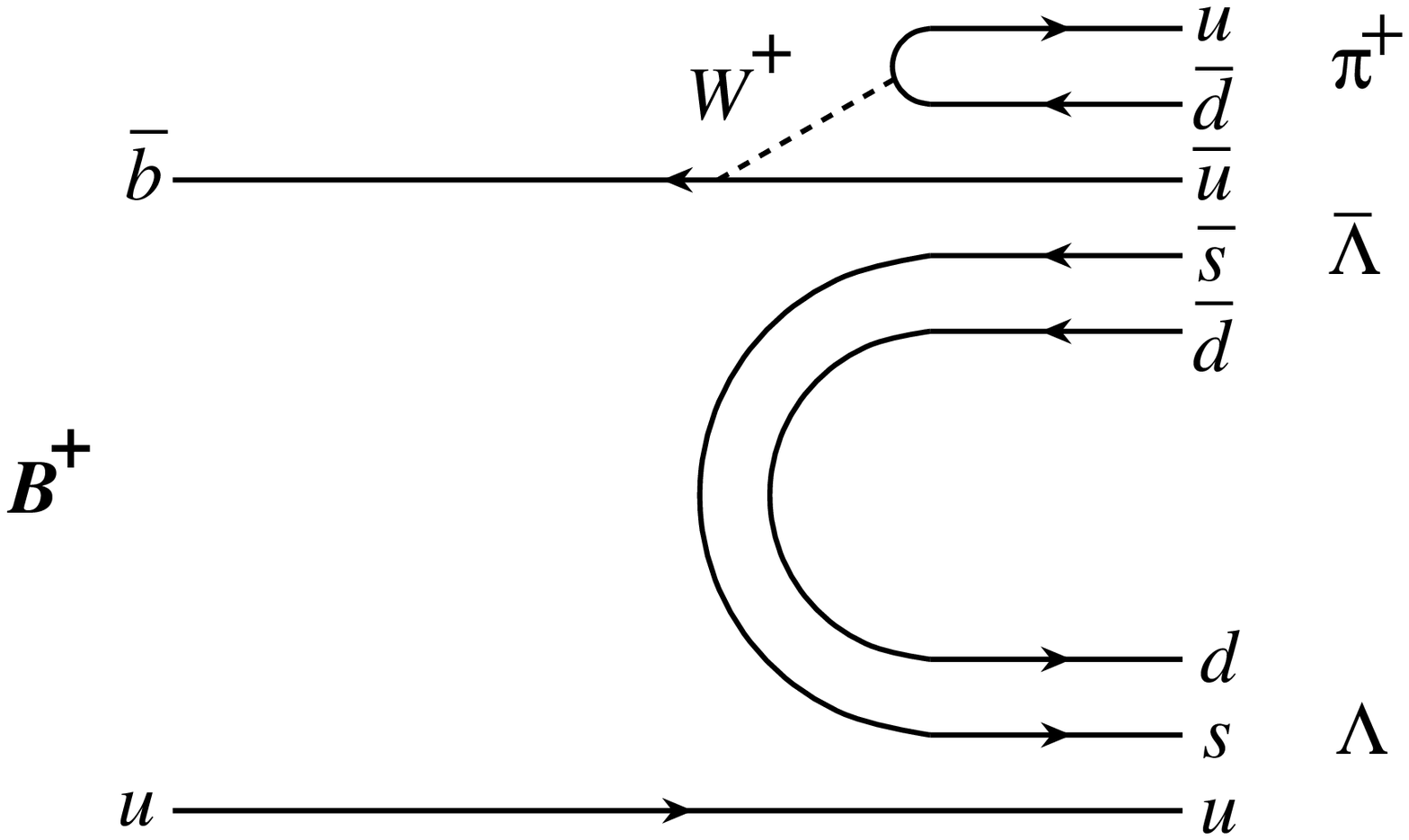}\\
\hskip -7.4cm {\bf (c)} \hskip 6.6cm {\bf (d)}\\
\vskip -1.8cm
\includegraphics[width=0.41\textwidth]{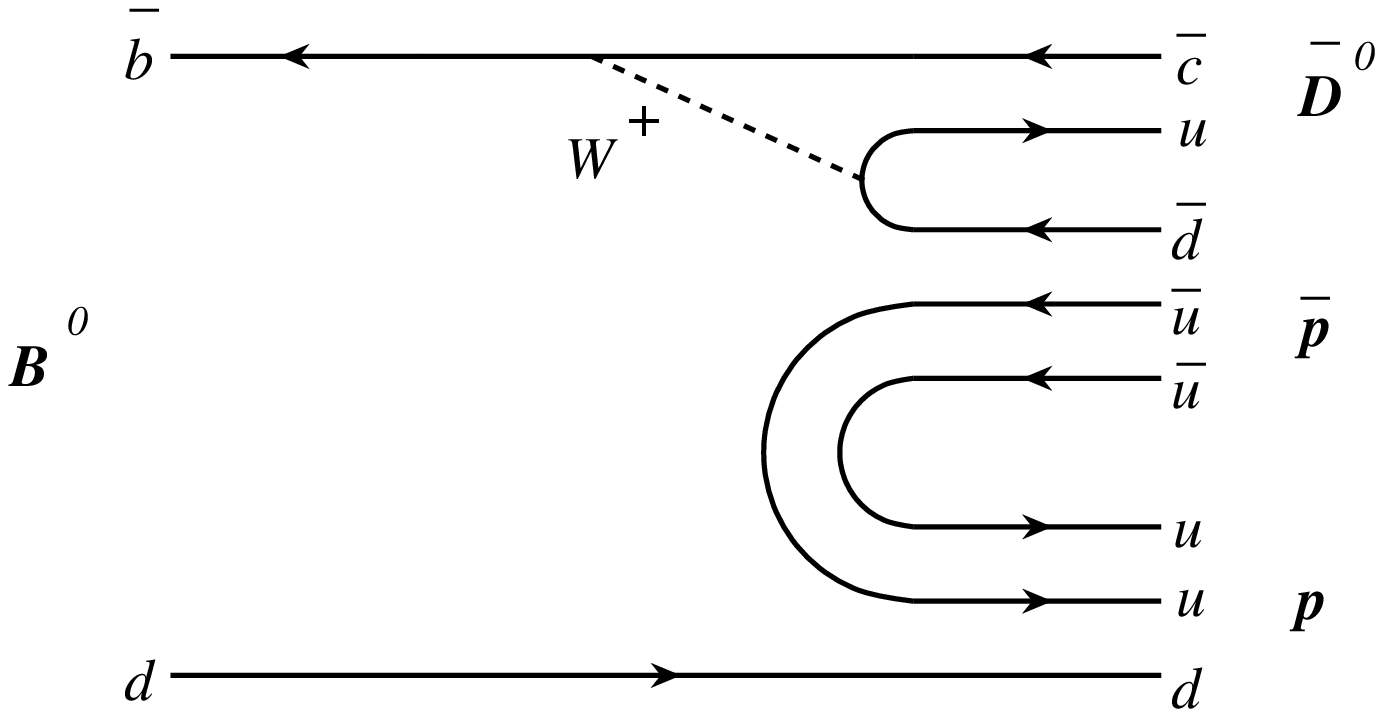}
\includegraphics[width=0.41\textwidth]{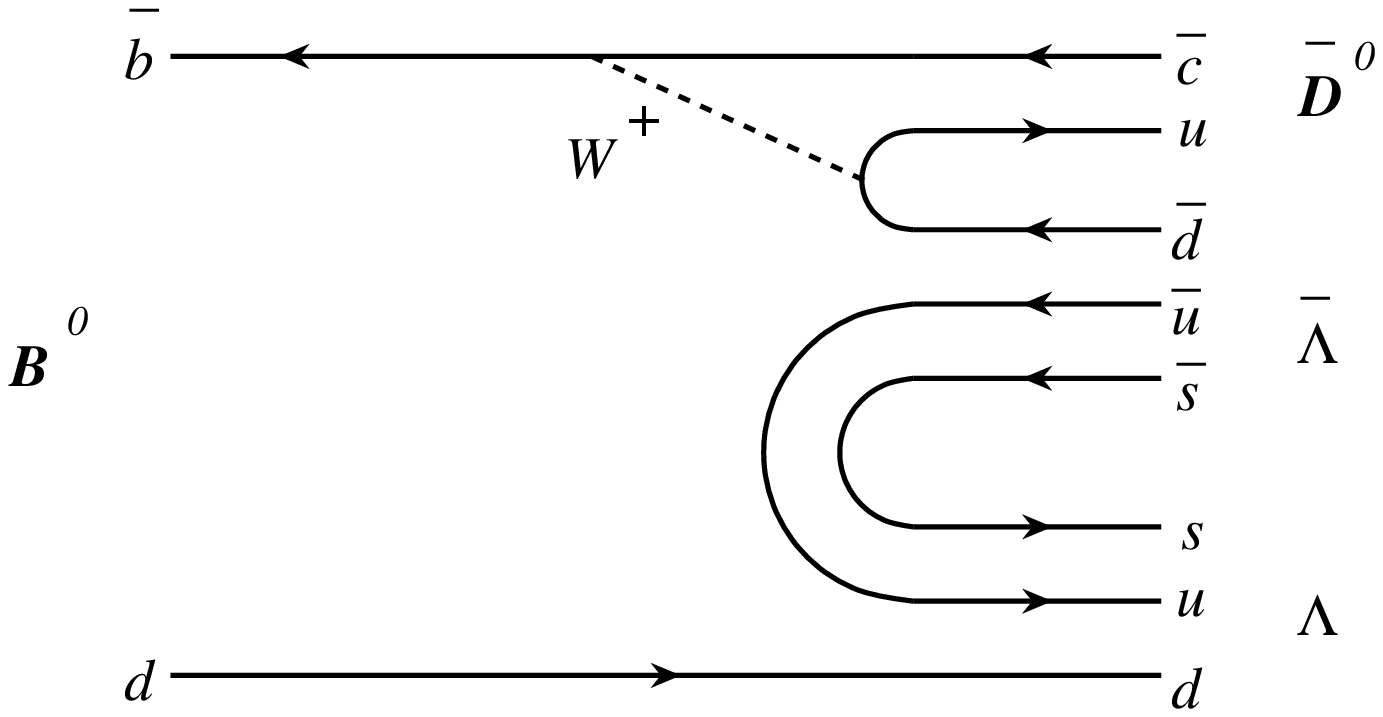}
\vskip -0.6cm
\caption{Possible diagrams that contribute to $\bp \to \pppi$/$\llpi$ and 
 $\bz \to \pp\bar{D^0}$/$\LL\bar{D^0}$.}
\label{fg:feyn2}
\end{center}
\end{figure}

 
  {\color{black}Our} results may indicate that the contribution of the $sd - \bar{s}\bar{d}$ popping diagram to
  $\bp \to \llpi$ shown in Fig.~\ref{fg:feyn2}(b), is suppressed relative to  
  the $ud - \bar{u}\bar{d}$ popping diagram shown in Fig.~\ref{fg:feyn2}(a).
In light of this observation,
we move to the $b \to c$ tree diagram (internal W emission) dominated decay 
$\bz \to \LL \bar{D^0}$ ($us-\bar{u}\bar{s}$ popping), {\color{black}which is} shown in Fig.~\ref{fg:feyn2}(d). 
We select the 1.852 GeV/c$^2 < M_{K^+\pi^-} < $ 1.877 GeV/c$^2$ region for $\bar{D^0}$
candidates and extract the $B$ yield. 
Figure~\ref{fg:lld0-fitting} shows the result of the fit. 
The signal yield is $5.53_{-2.35}^{+3.04}$ with a significance of 3.4$\sigma$.
The branching fraction 
{\color{black}
 $\mathcal{B} (\bz \to \LL\bar{D}^0)$
}
is $(1.05_{-0.44}^{+0.57})\times 10^{-5}$
{\color{black}
 $< 2.60 \times 10^{-5}$ at the 90\% confidence level. 
}
This branching fraction contrasts with the large, 
$(1.14 \pm 0.09) \times 10^{-4}$~\cite{PDG}, branching fraction observed for $\bz \to \pp\bar{D^0}$
($uu-\bar{u}\bar{u}$ popping) shown in Fig.~\ref{fg:feyn2}(c).
It appears that the diquark pair popping from the vacuum for $us - \bar{u}\bar{s}$ 
is considerably suppressed compared with $uu -\bar{u}\bar{u}$. 

\subsection{Comparison with predictions and previous measurements}
Table~\ref{tab:BR-comparison} shows a comparison of the branching fractions for 
$\bp \to \llk$, $\bp \to \llpi$ and $\bz \to \llkz$ decays 
to previous measurements~\cite{YJLee} and to theoretical predictions~\cite{prediction},
which are based on the experimental data for $B \to \pp h$ and $B \to p \bar{\Lambda} h$ decays. 

The branching fractions of $\bp \to \llk$ and $\bp \to \llpi$ seem to be consistent 
with the theoretical predictions and the results from previous measurements.

\begin{table}[htp]
 \caption{Comparison of branching fractions to previous results and theoretical predictions.}
  \label{tab:BR-comparison}
 \begin{center}
  \begin{tabular}{|c|ccc|}
   \hline
   Mode & Theoretical prediction\cite{prediction} & Previous measurement\cite{YJLee} & Results of our study \\
   \hline
   $\llk$~   & $(2.8 \pm 0.2) \times 10^{-6}$ & $(2.91^{+0.90}_{-0.70} \pm 0.38) \times 10^{-6}$ & $(3.38^{+0.41}_{-0.36} \pm 0.41) \times 10^{-6}$ \\
   $\llpi$~  & $(1.7 \pm 0.7) \times 10^{-7}$ & $< 2.8 \times 10^{-6}$ & ($< 0.94 \times 10^{-6}$)$^{\dag}$ \\
   $\LL\kz$~ & $(2.5 \pm 0.3) \times 10^{-6}$ &   -   & $(4.76^{+0.84}_{-0.68} \pm 0.61) \times 10^{-6}$ \\
   \hline
  \end{tabular}
 \end{center}
 \vskip -0.3cm
 \hspace*{140pt}$\dag$: This value is obtained in the threshold-mass-enhanced region.
\end{table}

\subsection{Systematic Study}
Systematic uncertainties are determined using high statistics control data samples.
\subsubsection{Reconstruction Efficiency} 
\begin{itemize}
\item Tracking uncertainty:\\
Tracking uncertainty is determined with
fully and partially reconstructed $D^*$ samples. 
It is about 1.3\% per charged track.
\item Particle identification uncertainty:\\
For proton identification, we use a  $\Lambda \to p \pi^-$ sample, 
while for $K/\pi$ identification we use a $D^{*+} \to D^0\pi^+$, $D^0 \to K^-\pi^+$ sample.
Note that the average efficiency difference for proton identification between data and MC 
has been corrected to obtain the final branching fraction measurements. 
The corrections are  7.41\%, 7.40\%, 7.40\%, 7.45\% and 7.48\% for the 
$\bp\to\llk$, $\bp\to\llpi$, $\bz\to\LL\kz$, $\bp\to\llkst$ and $\bz\to\llkstz(\bar{D^0})$ modes, respectively.
The uncertainties associated with
the particle identification corrections are estimated to be 2\% for the proton(anti-proton) from $\Lambda$($\bar{\Lambda}$) 
and 0.8\% for each kaon/pion identification. 
\item $\Lambda$ Reconstruction:\\
We vary the $\Lambda$ selection criteria to estimate
their impact on the systematic uncertainty. 
The uncertainties from the $\Lambda$ mass cut 
and requirements on kinematic variables 
are 1.9\% and 1.5\%, respectively.
For the reconstruction of $\Lambda$ and $\bar\Lambda$, we have an additional
uncertainty of 4.7\% in the
efficiency for displaced vertex reconstruction.
This is determined from the
difference between $\Lambda$ proper time distributions for data and MC
simulation. 
\item $\ks$ Reconstruction:\\
The uncertainty in $\ks$ reconstruction
is determined from a large sample of $D^- \to \ks\pi^-$ events.
We have an additional uncertainty of 4.9\% for $\ks$ reconstruction.
\item ${\mathcal R}$ selection:\\
We study the ${\mathcal R}$ continuum suppression
by varying the ${\mathcal R}$ cut value from 0 to 0.9 to check for a systematic trend.
\item Multiple Candidates:\\
{\label{MultiCount}}
{\color{black} The systematic uncertainty in the best $B$ candidate selection is determined 
by including multicandidate events satisfying the $\cal R$ cut value
when obtaining the signal yield and the efficiency for each mode. We then 
take the difference in the branching fractions with and without 
the best candidate selection as the systematic uncertainty.}
\item MC statistical uncertainty:\\
The MC statistical uncertainty is less than 2\%.
\end{itemize}
\subsubsection{Fitting Uncertainty}
\begin{itemize} 
\item PDF uncertainty:\\ 
A systematic uncertainty in the fit yield 
is determined by varying the parameters of the signal and background PDFs.
{\color{black} The assumption of uncorrelated PDFs for $\mb$ and $\de$ 
is studied by using 2D smoothed histogram
functions for both signal and $q\bar{q}$ MC events. The percentage change in the signal yield is about
0.8\%.}
{\color{black} According to our MC simulation study, 
the rare $B$ decays that will significantly affect our signal determination are 
$\bp \to \llk$ for $\bp \to \llpi$ mode and 
$B \to \Lambda \bar{\Sigma}^0$h for all $B \to \LL h$ modes.
The latter contributes a 0.5\% error and is included in the 
systematic error from fitting.
}
The uncertainty in the fit from the $M_{K\pi}$ PDF for continuum background is determined from 
the difference between the fit results for the $B \to \pp K^*$ modes using analytical functions
(a threshold function and a p-wave function) 
and using the smooth function obtained from the $M_{K\pi}$ distribution in sideband data~\cite{JHChen}.
We quote 1\% fitting uncertainties for the $M_{K\pi}$ PDF of continuum background
in $B \to \LL K^*$ modes.
We quote a 3.2\% fitting uncertainty for the $M_{K\pi}$ PDF of non-resonant $B \to \LL K\pi$, 
which is obtained from the difference in the fit results for the $\pp K\pi$ mode 
using an analytical function(the LASS function) and using a second order polynomial.
The second order polynomial is $(x-M_{K\pi(\rm lowerb)})\times(x-M_{K\pi(\rm upperb)})$,
where $M_{K\pi(\rm lowerb)}$ is 0.63325 GeV/$c^2$ and $M_{K\pi(\rm upperb)}$ is 1.8, 2.4 or 3.0 GeV/$c^2$.
The uncertainty in the fit for the $M_{K\pi}$ PDF of non-resonant $\pp K\pi$ is largest~(3.2\%)
when $M_{K\pi(\rm upperb)}$ is 1.8 GeV/$c^2$~\cite{JHChen}.
The total fitting uncertainties for 
$\bp\to\llk$, $\bp\to\llpi$, $\bz\to\llkz$, $\bz \to \llkstz$ and $\bp\to\llkst$ modes
are 1.5\%, 2.0\%, 1.5\%, 6.0\% and 6.0\%, respectively.
\item Fitting bias:\\
We use 800 simulated MC event sets to measure the {\color{black}difference} 
between the fit result and the expected value.
The bias is less than 1\% for both 2D and 3D fits.
\end{itemize}

\subsubsection{MC modeling} 
\begin{itemize}
\item Angular distribution of the proton in the $\Lambda$ rest frame:\\
{\color{black}
As described in ~\ref{sec:llk-costhetap},
the efficiency uncertainty due to the polarization of $\Lambda$ ($\bar{\Lambda}$)
is bypassed by using a model independent method based on data. However, to be conservative,
we quote the percentage difference between efficiencies obtained from the
model independent method and from the theoretically predicted $\cos\theta_{p}$ distribution~\cite{Suzuki}. 
This modeling uncertainty is about 4.3\%.}

\item Angular distribution of the $\bar{\Lambda}$ in the $\LL$ rest frame:\\
{\color{black} 
We choose the most significant mode, $\bp \to \llk$, with $\mll <$ 2.85 GeV/c$^2$  to obtain 
its $\cos\theta_{\bar{\Lambda}}$ \ref{sec:PolarAngleD} distribution, shown in Fig.~\ref{fg:llk-costhl}. 
Although it deviates significantly from a phase space distribution, 
the overall efficiency difference from a phase space MC sample is small 
since the efficiency versus $\cos\theta_{\bar{\Lambda}}$ 
is symmetric and flat. We assume that this effect is the same for all other decay modes.
Thus, the uncertainties from the MC modeling for the angular distribution of  
$\cos\theta_{\bar{\Lambda}}$ are determined to be 0.9\%.}

\item Angular distribution of kaon in $K^{*0}$ rest frame:\\
The uncertainties from the MC modeling of the $\cos\theta_K$ angular distribution 
in the $\llkstz$ mode about 2.5\%.
This value is determined from the difference between 
the efficiency in the threshold-mass-enhanced region
obtained from the $B$ yields using phase space MC event samples
and the efficiency calculated from the efficiency distribution function, 
the theoretical PDFs for the $K^{*0}$ meson and 
the ratio of the two helicity states obtained by fitting to data.
\end{itemize}

\subsubsection{Total systematic errors}
The systematic uncertainties for each decay channel are
summarized in Table~\ref{systematics}.
These uncertainties are summed in quadrature to determine the total systematic uncertainty for each mode.

\begin{table}[hp]
 \caption{Contributions to the systematic uncertainty(in \%).}
 \begin{center}
  \small
  \begin{tabular}{|c|cccccc|}
   \hline
   Source & $\llk$~ & $\llpi$~ & $\LL\kz$~ & $\llkst$~ & $\llkstz$~ & $\LL \bar{D^0}$~ \\
   \hline
   Tracking                     & 6.8 & 6.8 & 7.8 & 9.2 & 7.9 & 7.9 \\
   Proton ID                    & 4.0 & 4.0 & 4.0 & 4.0 & 4.0 & 4.0 \\
   Charged Kaon(Pion) ID        & 0.8 & 0.8 &  -  & 0.8 & 1.6 & 1.6 \\
   $\Lambda$ Reconstruction     & 4.7 & 4.7 & 4.7 & 4.7 & 4.7 & 4.7 \\
   $\Lambda$ Selection Cut      & 2.4 & 2.4 & 2.4 & 2.4 & 2.4 & 2.4 \\
   $\ks$ Reconstruction         &  -  &  -  & 4.9 & 4.9 &  -  &  -  \\
   $\mathcal R$ Selection       & 5.3 & 5.3 & 1.3 & 3.5 & 3.5 & 3.5 \\
   Multiple Candidates          & 0.9 & 0.9 & 2.4 & 1.4 & 1.4 & 1.4 \\
   MC Statistics                & 1.0 & 1.0 & 1.5 & 2.0 & 2.0 & 2.0 \\
   PDF uncertainties            & 1.7 & 2.2 & 1.7 & 6.1 & 6.1 & 3.7 \\
   Fitting bias                 & 1.0 & 1.0 & 1.0 & 1.0 & 1.0 & 1.0 \\   
   MC modeling                  & 4.4 & 4.4 & 4.4 & 4.4 & 5.1 & 4.4 \\
   Secondary decays~\cite{PDG}  & 1.6 &	1.6 &	1.6 &	1.6 &	1.6 &	2.0 \\
   Number of $B\bar{B}$ Pairs   & 1.4 & 1.4 & 1.4 & 1.4 & 1.4 & 1.4 \\
   \hline
   Total                        & 12.2 & 12.3 & 12.9 & 15.3 & 14.0 & 12.9 \\
   \hline
  \end{tabular}
 \end{center}
 \label{systematics}
\end{table}

\section{Summary} 
{\color{black}
Using 657 $ \times 10^6 B\bar{B}$ events, we 
observe  low mass $\mll$ enhancements near
threshold for both the $\llkz$ and $\llkstz$ modes, 
with 12.4$\sigma$ and 9.3$\sigma$ significance, respectively.
We update the branching fraction of $\bp \to \llk$ mode
superseding the previous measurement~\cite{YJLee}, 
and set upper limits on the modes $\bp \to \llkst$ and $\bp \to \llpi$ 
in the threshold-mass-enhanced region.
No significant signal is found in the related mode $\bz \to \LL\bar{D}^0$.
All the details are summarized in Table~\ref{summation-table}.
The small value of $\mathcal{B} (\bp \to \llpi)$,
the large value of $\mathcal{B} (\bz \to \llkz)$,
and the absence of a peaking feature in the $\cos\theta_{\bar{\Lambda}}$ distribution
for $\bp \to \llk$ indicate that the
underlying dynamics of $B \to \LL h$ are quite different from those of $B \to \pp h$. 
These results also imply that the $\bar{s}$ quark from $\bar{b} \to \bar{s}$ penguin diagram does not necessarily 
hadronize to form a $K^+$;
the probability of forming a $\bar{\Lambda}$ is not negligible.
In addition, because $\mathcal{B} (\bz \to \LL \bar{D^0} (\bp \to \llpi)$)
is much  smaller than $\mathcal{B} (\bz \to \pp \bar{D^0} (\bp \to \pppi$)),
it appears that diquark pair popping out from the vacuum for
$us - \bar{u}\bar{s}$  ($sd - \bar{s}\bar{d}$) is suppressed compared to
$uu -\bar{u}\bar{u}$ ($ud - \bar{u}\bar{d}$). 
}

\begin{table}[b]
 \caption{Summary of all $B \to \LL h$ results.}
 \begin{center}
  \begin{tabular}{|c|ccc|}
   \hline
   \multicolumn{4}{|c|}{Charmless branching fractions.}\\
   \hline
   Mode              & Yield & $\mathcal B(10^{-6})$ & Significances ($\sigma$)\\
   \hline
   $\bz \to \llkz$   & $49.1_{-7.1}^{+8.6}$ & $4.76_{-0.68}^{+0.84} \pm 0.61$ & 12.5\\
   $\bz \to \llkstz$ & $25.3_{-7.8}^{+9.4}$ & $2.46_{-0.72}^{+0.87} \pm 0.34$ & 9.0\\
   $\bp \to \llk$    & $103.4_{-11.2}^{+12.9}$ & $3.38_{-0.36}^{+0.41} \pm 0.41$ & 16.5\\
   \hline
   \multicolumn{4}{|c|}{Results in the threshold-mass-enhanced region.}\\
   \hline
   Mode              & Yield & $\mathcal B(10^{-6})$ & Significances ($\sigma$)\\
   \hline
   $\bp \to \llpi$   & $7.76_{-3.72}^{+4.49}$ & $< 0.94$ at 90\% C.L. & 2.5\\
   $\bp \to \llkst$  & $6.54_{-2.63}^{+3.37}$ & $2.19_{-0.88}^{+1.13} \pm 0.33$ & 3.7\\
                     &                        & ( $< 4.98$ at 90\% C.L. ) &  \\
   \hline
   \multicolumn{4}{|c|}{Related search.}\\
   \hline
   Mode              & Yield & $\mathcal B(10^{-5})$ & Significances ($\sigma$)\\
   \hline
   $\bz \to \LL \bar{D^0}$   & $5.53_{-2.35}^{+3.04}$ & $1.05_{-0.44}^{+0.57}\pm 0.14$ & 3.4\\
                             &                       & ( $< 2.60$ at 90\% C.L. ) &  \\
   \hline
  \end{tabular}
 \end{center}
 \vskip -0.3cm
 \label{summation-table}
\end{table}

We thank the KEKB group for the excellent operation of the
accelerator, the KEK cryogenics group for the efficient
operation of the solenoid, and the KEK computer group and
the National Institute of Informatics for valuable computing
and SINET3 network support. We acknowledge support from
the Ministry of Education, Culture, Sports, Science, and
Technology of Japan and the Japan Society for the Promotion
of Science; the Australian Research Council and the
Australian Department of Education, Science and Training;
the National Natural Science Foundation of China under
contract No.~10575109 and 10775142; the Department of
Science and Technology of India; 
the BK21 program of the Ministry of Education of Korea, 
the CHEP src program and Basic Research program (grant 
No. R01-2008-000-10477-0) of the 
Korea Science and Engineering Foundation;
the Polish State Committee for Scientific Research; 
the Ministry of Education and Science of the Russian
Federation and the Russian Federal Agency for Atomic Energy;
the Slovenian Research Agency;  the Swiss
National Science Foundation; the National Science Council
and the Ministry of Education of Taiwan; and the U.S. Department of Energy.

\clearpage


\begin{thebibliography}{99}
\bibitem{conjugate} {Throughout this report, inclusion of 
charge conjugate mode is always implied unless otherwise stated.}
\end{thebibliography}
\end{document}